\documentclass[graybox]{svmult}

\usepackage{mathptmx}       
\usepackage{helvet}         
\usepackage{courier}        
\usepackage{type1cm}        
\usepackage{makeidx}         
\usepackage{graphicx}        
\usepackage{multicol}        
\usepackage[bottom]{footmisc}

\newcommand\lsim{\mathrel{\rlap{\lower4pt\hbox{\hskip1pt$\sim$}}
        \raise1pt\hbox{$<$}}}
\newcommand\gsim{\mathrel{\rlap{\lower4pt\hbox{\hskip1pt$\sim$}}
        \raise1pt\hbox{$>$}}}
\newcommand{\msun}{{\rm M_{\odot}}}
\newcommand{\gt}{$>$}
\newcommand{\apj}{ApJ}
\newcommand{\aapr}{Astron. Astrophys. Rev.}
\newcommand{\jcap}{JCAP}
\newcommand{\prd}{Phys. Rev. D}
\newcommand{\aap}{A\&A}
\newcommand{\apjl}{ApJL}
\newcommand{\mnras}{MNRAS}
\newcommand{\aj}{AJ}
\newcommand{\na}{New Astronomy}
\newcommand{\apjs}{ApJS}
\newcommand{\nat}{{\it Nature}}
\newcommand{\araa}{ARA\&A}

\newcommand{\pasp}{PASP}

\newcommand{\Msol}{M_{\odot}}

\newcommand{\Mpc}{\;\mathrm{Mpc}}

\newcommand{\mh}{{M_{\bullet}}}

\newcommand{\HH}{H$_2$}

\makeindex             

\begin{document}

\title*{The Formation of the First Massive Black Holes}
\author{Zolt\'an Haiman}
\institute{Zolt\'an Haiman \at Department of Astronomy, Columbia University, 550 West 120th Street, New York, NY 10027 \email{zoltan@astro.columbia.edu}}
%
%
\maketitle

\abstract*{Supermassive black holes (SMBHs) are common in local
  galactic nuclei, and SMBHs as massive as several billion solar
  masses already exist at redshift $z=6$. These earliest SMBHs may
  grow by the combination of radiation--pressure--limited accretion
  and mergers of stellar-mass seed BHs, left behind by the first
  generation of metal-free stars, or may be formed by more rapid
  direct collapse of gas in rare special environments where dense gas
  can accumulate without first fragmenting into stars. This chapter
  offers a review of these two competing scenarios, as well as some
  more exotic alternative ideas.  It also briefly discusses how the
  different models may be distinguished in the future by observations
  with {\it JWST, LISA} and other instruments.}

\abstract{Supermassive black holes (SMBHs) are common in local
  galactic nuclei, and SMBHs as massive as several billion solar
  masses already exist at redshift $z=6$. These earliest SMBHs may
  grow by the combination of radiation--pressure--limited accretion
  and mergers of stellar-mass seed BHs, left behind by the first
  generation of metal-free stars, or may be formed by more rapid
  direct collapse of gas in rare special environments where dense gas
  can accumulate without first fragmenting into stars. This chapter
  offers a review of these two competing scenarios, as well as some
  more exotic alternative ideas.  It also briefly discusses how the
  different models may be distinguished in the future by observations
  with {\it JWST, LISA} and other instruments.}

\section{Introduction}
\label{sec:Intro}

The discovery of about two dozen bright quasars with luminosities
$\gsim 10^{47}~{\rm erg~s^{-1}}$ at redshift $z\simeq 6$ suggests that
some supermassive black (SMBHs) as massive as a few$\times10^9~\msun$
have been already assembled when the universe was less than 1 Gyr old
(see, e.g., ref.~\cite{Fanreview06} for a review).  These
high-redshift quasars are exceedingly rare, with a space density of
order $\sim1{\rm Gpc^{-3}}$, and can only be found in large surveys of
the sky, such as the Sloan Digital Sky Survey (SDSS), or the
smaller--area but deeper CFHQS~\cite{Willott+10a} and
UKIDSS~\cite{Lawrence+07} surveys.  These quasars overall appear to be
``fully developed'', with spectra and metallicity patterns that appear
remarkably similar to their counterparts at moderate
redshifts~\cite{Fan+03}.  Indeed, if one selects individual quasars
with the same luminosity, their properties show little evolution with
cosmic epoch.\footnote{A few possibly important exceptions to this are
discussed in \S~\ref{zoltansubsec:sdss}.}  This implies that the
behavior of individual quasars is probably determined by local physics
near the SMBH and is not directly coupled to the cosmological context
in which the SMBH is embedded.  However, it is clear that the quasar
population as a whole does evolve over cosmic timescales.
Observations from $0\lsim z \lsim 6$ in the optical (e.g., the
Anglo-Australian Telescope's Two Degree Field, or 2dF, and the Sloan
Digital Sky Survey, or SDSS) and radio bands~\cite{Shaver+96} show a
pronounced peak in the abundance of bright quasars at $z\approx 2.5$.
A similar behavior has been confirmed in X-ray
observations~\cite{Silverman+08}.

The cosmic evolution of quasar black holes between $0 \lsim z \lsim 6$
is likely driven by a mechanism other than local physics near the
hole.  This is reinforced by the fact that the timescale of activity
of individual quasars is significantly shorter than cosmic timescales
at $z\lsim 6$, both on theoretical grounds ($\sim 4\times10^7$ yr, the
e-folding time for the growth of mass in a SMBH, whose accretion
converts mass to radiation with an efficiency of $\epsilon=\dot M
c^2/L_{\rm Edd}\sim 10\%$) and is limited by its own [Eddington]
luminosity), and using the duty cycle of quasar activity inferred from
various observations (also $\sim 10^7$ yr but with large
uncertainties, e.g.~\cite{Martini04} and references therein; see also
\cite{HCO04,Shankarreview,SWM09,SWS10}).

In the cosmological context, it is tempting to link the evolution of
massive quasar black holes with that of dark matter (DM) halos
condensing in a Cold Dark Matter (CDM) dominated universe, as the halo
population naturally evolves on cosmic timescales~\cite{ER88}.
Indeed, this connection has proven enormously fruitful and has
resulted in the following broad picture: the first massive
astrophysical black holes appear at high redshifts ($z\gsim 10$) in
the shallow potential wells of low mass ($\lsim 10^8~\msun$) dark
matter halos. These black holes grow by mergers and gas accretion,
evolve into the population of bright quasars observed at lower
redshifts, and eventually leave the SMBH remnants that are ubiquitous
at the centers of galaxies in the nearby universe.

In this picture, the presence of few$\times10^9~\msun$ SMBHs at $z>6$
presents a puzzle~\cite{HL01}\footnote{More generally, the non-trivial
cosmological implications of the existence of massive BHs at early
times was noted already when quasars were first found at redshifts
$z>4$~\cite{Turner91}.}.  Metal--free stars, with masses $\sim
100~\msun$, are expected to form at redshifts as high as $z\gsim 25$
\cite{ABN02,BCL02,YOH08}, and leave behind remnant BHs with similar
masses \cite{CBA84,Heger+03}.  However, the natural time-scale,
i.e. the Eddington time, for growing these seed BHs by $\gsim 7$
orders of magnitude in mass is comparable to the age of the universe
(e.g. ref.~\cite{HL01}). This makes it difficult to reach $10^9~\msun$
without a phase of rapid (at least modestly super--Eddington)
accretion, unless a list of optimistic assumptions are made in
hierarchical merger models, in which multiple seed BHs are allowed to
grow without interruption, and to combine into a single SMBH
\cite{Haiman04a,YM04,BSF04,Shapiro05,VR06,PDC07,Li+07,SSH09,TH09}.

An alternative class of explanations involves yet more rapid gas
accretion or collapse
\cite{OH02,BL03,KBD04,LN06,SS06,BVR06,VLN08,WA08,RH09b,SSG10,SBH10}. In
this family of models, primordial gas collapses rapidly into a SMBH as
massive as $10^4-10^6~\msun$, either directly, or possibly by
accreting onto a pre--existing smaller seed BH \cite{VR05}, or going
through the intermediate state of a very massive star \cite{BL03}, a
``quasistar'' \cite{BRA08}, or a dense stellar cluster
\cite{OSH08,DV09}.  These so--called ``direct collapse'' models
involve metal--free gas in relatively massive ($\gsim 10^8~\msun$)
dark matter halos at redshift $z\gsim 10$, with virial temperatures
$T_{\rm vir}\gsim 10^4$K.  The gas that cools and collapses in these
halos must avoid fragmentation, shed angular momentum efficiently, and
collapse rapidly.

Many uncertainties about each of the above scenarios remain, and the
astrophysical process(es) responsible for the formation of the
earliest massive black holes (and indeed for the presence of SMBHs at
all redshifts) remain poorly understood.  In this review, we focus on
the emergence of the first generation of black holes, though many of
the important questions are quite general and apply equally to
subsequent generations of black holes.  This review is organized as
follows. 
In \S~\ref{sec:first}, we describe theoretical expectations for the
formation and growth of these black holes within the paradigm of
hierarchical CDM cosmologies.
In \S~\ref{sec:theory}, we
``zoom in'' and consider the local physics of black hole formation,
and various pathways which could lead to the early presence of supermassive black holes.
In \S~\ref{sec:observations}, we summarize several relevant recent
observations that have implications for early black holes, and
speculate on the power of future observations to probe the physics of
the first BHs. We offer our conclusions in \S~\ref{sec:conclude}.

This chapter is an expanded an updated version of an earlier
review~\cite{HQ04}.  Another recent, complimentary review on SMBH
formation at high redshift can be found in \cite{Volonteri10}.

\section{First Structure Formation}
\label{sec:first}

In this section, we sketch some basic theoretical arguments relevant
to the formation of structure in the early universe. We then discuss
formation mechanisms for SMBHs.

\subsection{Cosmological Perturbations as the Sites of the First Black Holes}
\label{zoltansubsec:cosmo}

Measurements of the Cosmic Microwave Background (CMB) anisotropies by
the {\em Wilkinson Microwave Anisotropy Probe (WMAP)}, determinations
of the luminosity distance to distant type Ia Supernovae, and other
observations have led to the emergence of a robust ``best fit''
cosmological model with energy densities in CDM and ``dark energy'' of
$(\Omega_{\rm M},\Omega_{\rm \Lambda})\approx (0.3,0.7)$ (see, e.g.
~\cite{WMAP7}, for the seven-year {\em WMAP} results, and its
combination with other datasets).

The growth of density fluctuations and their evolution into nonlinear
dark matter structures can be followed in this cosmological model from
first principles by semi-analytic methods~\cite{PS74,SMT01}.  More
recently, it has become possible to derive accurate dark matter halo
mass functions directly in large cosmological N-body
simulations~\cite{Jenkins+01}, with different codes agreeing at the
10\% level, and mass functions measured down masses as low as $\sim
10^6~{\rm M_\odot}$ and high redshifts as high as $z\approx 30$
(e.g.~\cite{Lukic+07,Reed+07}).

Within the $\Lambda$CDM model, with a scale--invariant primordial
power spectrum, robust predictions can therefore be made for the dark
matter halos.  Structure formation in such a universe is
``bottom-up'', with low--mass halos condensing first.  Halos with the
masses of globular clusters ($10^{5-6}~\msun$) are predicted to have
condensed from $\sim 3\sigma$ peaks of the initial primordial density
field as early as $\sim1\%$ of the current age of the universe, or at
redshifts of $z\sim 25$.  These predictions are limited mainly by the
$5-10\%$ uncertainty in the normalization of the primordial power
spectrum, $\sigma_{8}$, and by the need to extrapolate the
power--spectrum 2-3 orders of magnitude below the scales on which it
has been directly constrained.  In warm dark matter models, with
particle masses of order $\sim1$keV or less, free--streaming would
result in a sharp exponential suppression of the fluctuation power on
the relevant scales (masses below $10^8~{\rm M_\odot}$), and could
significantly reduce the number of DM halos at the earliest
redshifts~\cite{BHO01,Yoshida+03b}.

It is natural to identify the first collapsed DM halos as the sites
where the first astrophysical objects, including the first black
holes, were born.  The nature of the objects that form in these early
dark matter halos is currently one of the most rapidly evolving
research frontiers in astronomy.

\subsection{Chemistry and Gas Cooling at High Redshifts}
\label{zoltansubsec:H2}

Baryonic gas that falls into the earliest nonlinear dark matter halos
is unable to cool efficiently, and is shock heated to the
characteristic virial temperatures less than a few hundred Kelvin. It
has long been pointed out~\cite{Binney77,RO77,WR78} that such gas
needs to lose its thermal energy efficiently (within about a dynamical
time) in order to continue contracting, or in order to fragment. In
the absence of any dissipation, it would simply reach hydrostatic
equilibrium and would eventually be incorporated into a more massive
halo further down the halo merger hierarchy.  While the formation of
nonlinear dark matter halos can be followed from first principles, the
cooling and contraction of the baryons, and the ultimate formation of
stars or black holes in these halos, is much more difficult to model
ab initio.

The gas content of a cosmological perturbation can contract together
with the dark matter only in dark halos above the cosmological Jeans
mass, $M_{\rm J}\approx 10^4~\msun [(1+z)/11]^{3/2}$, in which the
gravity of dark matter can overwhelm thermal gas pressure.  Recent
work~\cite{TH10} has shown that immediately following recombination
(at redshift $z\sim 1,000$), the baryons develop coherent streaming
motions relative to the dark matter, with relative speeds of order
$30~{\rm km~s^{-1}}$, on scales of a few Mpc.  These relative
velocities decay as $\propto(1+z)$, and reduce to $\sim 1~{\rm
km~s^{-1}}$ by $z\sim 30$, comparable to the velocity dispersions in
the smallest dark matter halos at this epoch.  In the somewhat more
massive halos in which the baryons can typically cool efficiently
(with velocity dispersions of several ${\rm km~s^{-1}}$ at $z\lsim
30$; see below), the streaming motions are expected to have at most a
modest effect, reducing the gas fraction within the virial radius by a
factor of $\sim$two~\cite{TBH11,Maio+11,Stacy+11,Greif+11b,DPS11}.
However, as we will see below, the ``stellar-seed'' model for SMBH
growth relies on one (or a few) ``special'' rare seed BHs that form in
few $\times\sim 10^5~{\rm M_\odot}$ halos at redshifts as high as
$z\sim 30$. This scenario appears vulnerable to the streaming motions.

In the earliest, chemically pristine clouds, radiative cooling is
dominated by ${\rm H_2}$ molecules. As a result, gas phase ${\rm H_2}$
``astro-chemistry'' is likely to determine the epoch when the first
stars and black holes appear (primordial molecular chemistry, focusing
on the role of ${\rm H_2}$ early structure formation was reviewed
by~\cite{AH01}).  Several papers have constructed complete gas-phase
reaction networks and identified the two possible ways of gas-phase
formation of ${\rm H_2}$ via the ${\rm H^-}$ or ${\rm H_2^+}$
channels.  These were applied to derive the ${\rm H_2}$ abundance
under densities and temperatures expected in collapsing high redshift
objects~\cite{Hirasawa69,Matsuda69,PSS83,LS84,SK87,Kang+90,KS92,SGB94}.
Studies that incorporate ${\rm H_2}$ chemistry into cosmological
models and that address issues such as non-equilibrium chemistry,
dynamics, or radiative transfer have appeared relatively more
recently. Ref~\cite{HTL96} used spherically symmetric simulations to
study the masses and redshifts of the earliest objects that can
collapse and cool via ${\rm H_2}$; their findings were confirmed by a
semi-analytic treatment~\cite{Tegmark+97}.  The first fully three
dimensional cosmological simulations that incorporate ${\rm H_2}$
chemistry and cooling date back to refs~\cite{OG96,GO97} and
\cite{Abel+97}.

The basic picture that emerged from these papers is as follows.  The
${\rm H_2}$ fraction after recombination in the smooth
``protogalactic'' gas is small ($x_{\rm H2}=n_{\rm H2}/n_{\rm H}\sim
10^{-6}$). At high redshifts ($z~\gsim 100$), ${\rm H_2}$ formation is
inhibited, even in overdense regions, because the required
intermediaries ${\rm H_2^+}$ and H$^-$ are dissociated by cosmic
``microwave'' background (CMB, but with the typical wavelength then in
the infrared) photons.  However, at lower redshifts, when the CMB
photons redshift to lower energies, the intermediaries survive, and a
sufficiently large ${\rm H_2}$ abundance builds up inside collapsed
clouds ($x_{\rm H2}\sim 10^{-3}$) at redshifts $z~\lsim 100$ to cause
cooling on a timescale shorter than the dynamical time.  Sufficient
\HH\ formation and cooling is, however, possible only if the gas
reaches temperatures in excess of $\sim 200$~K or masses of a few
$\times~10^5~\msun [(1+z)/11]^{-3/2}$ (note that while the
cosmological Jeans mass increases with redshift, the mass
corresponding to the cooling threshold, which is well approximated by
a fixed virial temperature, has the opposite behavior and decreases at
high redshift).  The efficient gas cooling in these halos suggests
that the first nonlinear objects in the universe were born inside
$\sim 10^5~\msun$ dark matter halos at redshifts of $z\sim 20-30$,
corresponding to an $\sim3-4\sigma$ peaks of the primordial density
peak (of course, yet rarer low-mass halos exist even earlier - the
first one within our Hubble volume collapsing as early as $z\approx
60$~\cite{NNB06}).

The behavior of metal-free gas in such a cosmological ``minihalo'' is
a well posed problem that has been addressed in three dimensional
numerical simulations.  The first series of such
simulations~\cite{ABN00,ABN02,BCL99,BCL02,Yoshida+03a} were able to
follow the contraction of gas to much higher densities than preceding
studies. They have shown convergence toward a temperature/density
regime of ${\rm T \sim 200~{\rm K}}$, ${\rm n \sim 10^{4}~{\rm
cm}^{-3}}$, dictated by the critical density at which the excited
states of ${\rm H_2}$ reach equilibrium and cooling becomes less
efficient~\cite{GP98}.  These simulations suggested that the gas does
not fragment further into clumps below sizes of $10^{2}-10^{3}~\msun$,
but rather it forms unusually massive stars.  Very recent simulations
reached higher resolution than the earlier ones, and, in some cases,
using sink particles, were able to continue their runs beyond the
point at which the first ultra-dense clump
develops~\cite{Turk+09,Stacy+10,Greif+11a,Prieto+11}.  These
simulations suggest that at least in some cases, the gas in the
central regions does, eventually, fragment into two or more distinct
clumps, raising the possibility that the first stars formed in pairs,
or even in higher--multiple systems.

The masses of the first stars would then presumably be reduced.  The
initial mass function (IMF) of the first stars is crucial, and is
indeed one of the most important uncertainties for early BH formation.
This is because massive stars would naturally leave behind black hole
seeds, which can subsequently grow by mergers and accretion into the
SMBHs.  Interestingly, massive stars appear to have an ``either/or''
behavior. Non-rotating stars with masses between $\sim 40-140~\msun$
and above $\sim 260~\msun$ collapse directly into a black hole without
an explosion, and hence without ejecting their metal yields into the
surrounding medium, whereas stars in the range $\sim 140-260~\msun$
explode as pair--instability supernovae without leaving a
remnant~\cite{Heger+03}.  In contrast, stars with initial masses $\sim
25-40~{\rm M_\odot}$ still leave BH remnants but also eject metals,
whereas those with masses $M\lsim 25~{\rm M_\odot}$ do not leave any
BH remnants.  This dichotomy is especially interesting because early
massive stars are attractive candidates for polluting the IGM with
metals at high redshifts~\cite{MFR01,WQ00}.  It is likely that the
first stars had a range of masses, in which case they could contribute
to both metal enrichment and to the seed black hole population, with a
relative fraction that depends sensitively on their initial mass
function (IMF).

\section{Massive Black Hole Formation}
\label{sec:theory}

Having reviewed the general problem of structure formation at high
redshifts, we now focus on the question of how the first SMBHs were
assembled.  It is worth emphasizing that this is an unsolved problem
-- indeed, it is not entirely clear even whether the first nonlinear
objects in the universe were stars or black holes, and whether
galaxies or their central black holes formed first~\cite{Haiman04c}.
The leading ideas related to the formation of SMBHs at high redshifts
can be broadly divided into three areas: (1) formation of seed black
holes from ``normal'' stellar evolution and subsequent
Eddington-limited accretion, (2) rapid direct collapse of gas to a
SMBH, usually via a supermassive star/disk, and (3) formation of a
SMBH (or an IMBH seed) by stellar dynamical processes in dense stellar
systems, such as star clusters or galactic nuclei.  It is, of course,
possible that all of these processes could be
relevant~\cite{BR78,Rees84}.

\subsection{Growth from Stellar-Mass Seeds}
\label{sec:stellarseed}

\subsubsection{Basic Ingredients and Uncertainties}
\label{sec:stellarseeduncertainties}

Perhaps the most natural possibility is that early SMBHs grow by gas
accretion out of stellar--mass seed black holes, left behind by early
generation of massive stars.  If the subsequent gas accretion obeys
the Eddington limit and the hole shines with a radiative efficiency of
10\%, then the time it takes for a SMBH to grow to the size of
$3\times 10^9~\msun$ from a stellar seed of $\sim 100~\msun$ is
$3\times 10^7~{\rm ln}(3\times10^9/100)~{\rm yr}\sim 7\times 10^8~{\rm
yr}$.  This is comparable to the age of the universe at $z=6$ ($\sim
9\times 10^8~{\rm yr}$ for a flat $\Lambda$CDM universe with
$H_0=70~{\rm km~s^{-1}~Mpc^{-1}}$ and $\Omega_M=0.3$).  Therefore, the
presence of these SMBHs is consistent with the simplest model for
black hole growth, provided that {\em (i) the seeds are present early
on (at $z\gsim 15$; see below), (ii) and the near-Eddington growth is
uninterrupted}.  As the $\sim 10^{5-6}~{\rm M_\odot}$ host halo of the
initial seed BH gets incorporated into the $\sim 10^{12-13}~{\rm
M_\odot}$ host halo of the $z\approx 6$ SMBH, it grows by $\sim7$
orders in magnitude, and experiences a large number of mergers with
other, comparable-sized, halos. These merger partners may (or may not)
have a growing BH at their centers.  Therefore, these mergers need to
be taken into account, and the ``stellar seed'' model most likely can
not be viewed as that of a single seed BH, growing in isolation.

Several authors have worked out the growth of SMBHs from stellar--mass
seeds, by following the build-up of dark DM halos, and using simple
prescriptions to track the formation of seed BHs, their subsequent
growth by accretion, and their mergers.  This can be done either
semi--analytically \cite{HL01,WL03a,Haiman04a,Shapiro05}, using
Monte-Carlo realizations of the DM merger trees
\cite{YM04,BSF04,VR06,TH09}, or based on cosmological hydrodynamics
simulations \cite{Li+07,PDC07,SSH09}.  As noted in the Introduction,
the uncertainties about the statistics of the DM halo merger trees are
essentially negligible\footnote{At least in principle, since halo mass
functions in large N-body simulations agree at the few percent level.
In practice, however, there can be significant disagreements between
Monte-Carlo halo merger trees made with different
algorithms~\cite{ZFM08}.}, since DM halo formation has been directly
resolved in numerical simulations at the relevant low masses (down to
$\sim 10^6~{\rm M_\odot}$) and high redshifts (out to $z\approx 30$).
The most important -- and still highly uncertain -- ingredients of
this 'stellar seed' scenario can be summarized as follows.

{\em (i) What is the smallest possible mass (or virial temperature,
$T_{\rm seed}$) for early DM halos in which PopIII stars can form?} A
reasonable answer is $T_{\rm seed}=$few $\times$ 100 K, which allows
molecular ${\rm H_2}$--cooling \cite{HTL96,Tegmark+97}.

{\em (ii) In what fraction ($f_{\rm seed}$) of these halos do seed BHs
actually form?}  This is a much more difficult question, since various
feedback processes (due to radiation, metal pollution, or mechanical
energy deposition) could suppress PopIII star formation in the vast
majority of early low--mass halos (~\cite{HAR00,HaimanHolder03}; see
also a recent review \cite{BFH09}).  Interestingly, the {\it WMAP}
measurement of the electron scattering optical depth provides
empirical evidence that such negative feedback took place early on and
shaped the reionization history \cite{HB06}.  The answer also depends
on the IMF of PopIII stars, since, as noted above, whether the stars
leave a BH remnant or explode as pair instability SNe depends on their
masses. The dividing mass, $\approx 25~{\rm M_\odot}$, was evaluated
in non-rotating stellar evolution models~\cite{Heger+03}, whereas
recent simulations indicate that the first stars in minihalos have
significant rotation~\cite{SBL11}.  Rotation can help drive winds and
prevent BH formation entirely, or can produce a hypernova and reduce
the mass of the remnant BH. Finally the velocity dispersions of the
lowest-mass minihalos are only a few ${\rm km~s^{-1}}$, only a factor
of few higher than the residual bulk streaming motions between the gas
and the DM halo at $z\gsim 20$. These streaming motions can therefore
reduce the gas fractions in the earliest minihalos and also lower the
stellar masses by driving turbulence~\cite{Greif+11b}.

{\em (iii) What is the time--averaged accretion rate of the seed BHs?}
This is conveniently parameterized by a duty cycle $f_{\rm duty}$,
defined as the fraction of the mass accretion rate that would produce
the Eddington luminosity, if $\epsilon\approx 10\%$ of the rest mass
was converted to radiation (so that $f_{\rm duty}=1$ is the fiducial
Eddington rate).  Radiative feedback is usually expected to lead to
sub--Eddington rates (e.g. \cite{AWA09}), and in spherical symmetry,
the accretion was recently to shown to be episodic, with $f_{\rm
duty}\approx 0.3$ \cite{Milos+09}. The expectation is therefore that
$f_{\rm duty}$ is less than unity. In practice, if the accretion is
radiatively inefficient, or if the radiation is trapped or is beamed
and ``leaks out'', then $f_{\rm duty}$ could exceed unity (see more on this below).  

{\em (iv) Finally, what happens when DM halos merge?}  The simplest
and most optimistic assumption is that the BHs promptly coalesce, as
well.  However, even if dynamical friction on the DM (and on any stars
present in later stages of the merger hierarchy) is efficient, it is
possible that, due to the radiation of its parent star, the remnant
BHs are no longer embedded in dense enough gas to allow
this. Furthermore, even if the BHs coalesce, the merged binary BH can
suffer strong gravitational recoil at the time of the merger, due to
the linear momentum carried by the anisotropic emission of
gravitational waves (e.g. \cite{Baker+08} and references therein).
Such a ``kick'' can eject the BH from the shallow potential wells
($\sim 1$km/s) of the early halos, and the BH will be effectively
lost. While kicks for comparable-mass BHs with random spins are of the
order of $\sim 100$ km/s, the kick speed depends strongly on the mass
ratio and on the spin vectors of the two BHs.  In particular, kicks
become very small ($\lsim 1$km/s) for mass ratios $q\equiv
M_1/M_2\lsim 10^{-2}$, irrespective of BH
spins~(e.g.~\cite{Baker+08}). This may be key to avoid loosing growing
seed BHs by ejection, and thus for the buildup of SMBHs early on.

\begin{figure}[t]
  \includegraphics[height=.6\textheight]{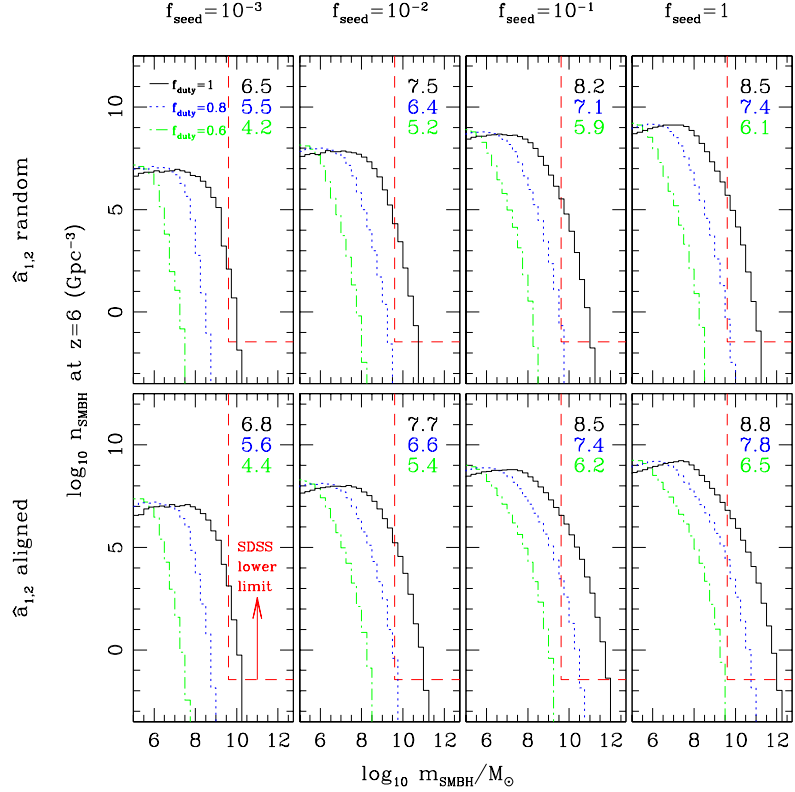}
  \caption{The comoving number densities of SMBHs in different mass
    bins at redshift $z=6$.  The 24 different models shown in the
    figure assume different parameter combinations as follows. The
    columns, from left to right, adopt $f_{\rm seed}=10^{-3}$,
    $10^{-2}$, $10^{-1}$, $1$.  The top row assumes a random binary
    spin orientation, and the bottom row assumes that BH spins are
    aligned with the binary's orbital angular momentum.  In each
    panel, the time--averaged mass--accretion rates, in Eddington
    units, are assumed to be either $f_{\rm duty}=1$ (black solid
    curves), $f_{\rm duty}=0.8$ (blue dotted), and $f_{\rm duty}=0.6$
    (green dash-dotted).  The numbers in the upper-right corners
    represent the total mass density in SMBHs
    $\log_{10}[\rho_{\bullet}/(\Msol \Mpc^{-3})]$ for each model.  The
    red dashed line demarcates the abundance of $z\approx 6$ SMBHs
    with $m\gsim 10^{9.6}\Msol$ already observed in the SDSS (adapted
    from ref.~\cite{TH09}).}
\label{fig:SMBHmassfunction}
\end{figure}

\subsubsection{Worked Illustrative Examples}
\label{sec:stellarseedexample}

In Figure~\ref{fig:SMBHmassfunction}, we show SMBH mass functions at
$z=6$, illustrating the impact of three of the most uncertain basic
assumptions above, taken from a recent example of the Monte Carlo
merger tree approach \cite{TH09}.  The mass functions were constructed
from the merger histories of $\approx10^5$ DM halos with masses
$M>10^{8}\Msol$ at redshift $z=6$.  The upper right region in each
panel, demarcated by the red dashed rectangle, shows the observational
constraint on the SMBH space density.  Each galaxy was modeled with a
spherically symmetric mass distribution consisting of a DM halo with a
Navarro-Frenk-White (NFW) profile~\cite{NFW97}, and a more cuspy
baryonic component (with $\rho\propto r^{-2.2}$, suggested by 3D
simulations).  At the time of a merger, the trajectories of kicked BH
-- ejections, or oscillations damped by dynamical friction -- were
followed explicitly by one-dimensional orbital calculations.

Figure~\ref{fig:SMBHmassfunction} shows that a robust conclusion for a
model to produce enough $z=6$ SMBHs is that $f_{\rm duty}\gsim 0.6$ --
namely the $\approx 100~{\rm M_\odot}$ stellar seed BHs must accrete
near the Eddington rate nearly all the time.  (Note that this value is
excluded in the spherically symmetric case \cite{MCB09}.) The initial
BH occupation fraction also has to be $f_{\rm seed}\gsim 10^{-3}$.
Another interesting, and less intuitive conclusion, is that if the
initial seeds are rare ($f_{\rm seed}=10^{-3}-10^{-2}$), then
gravitational kicks do {\em not} have a big impact, and it makes
little difference to the SMBH mass function whether spins are aligned
or randomly oriented (this can be seen by comparing the bottom and top
panels in Fig.~\ref{fig:SMBHmassfunction}).  This is because the few
``lucky'' seeds that form earliest (at $z\gsim 25$) have a chance to
grow by $\gsim$ two orders of magnitude in mass before encountering
their first merger.  The masses of the two BHs at this first merger
are then very unequal ($q=M_1/M_2\lsim 0.01$), making kick velocities
too low to lead to ejection.  It is important to emphasize, however,
that the model trajectories for the kicked BHs assume spherical
symmetry~\cite{MQ04,BL08,GM08,TH09}. In gas-rich galaxies, most of the
dynamical friction occurs due to the dense baryons at the center of
the potential~\cite{TH09,Guedes+11}. In asymmetric potentials, the
kicked BH does not return to the central region - its oscillations are
damped less quickly~\cite{Guedes+09} and the accretion rate onto the
oscillating hole is also suppressed.

An important additional issue is that in those models that satisfy the
observational constraint on the SMBH abundance, the massive end of the
SMBH mass function is extremely steep.  This prediction is not
surprising, as the most massive SMBHs reside in few $\times10^{12}{\rm
M_\odot}$ halos, which probe the $5\sigma$ tail of the halo mass
function at $z=6$ (and there are indeed $\approx 10^8$ (!) times as
many few$\times10^{9}{\rm M_\odot}$ halos, which host $\sim 10^6{\rm
M_\odot}$ BHs).  As a result, the total mass density in SMBHs with
masses above $\gsim 10^5{\rm M_\odot}$ BHs (shown by the numbers in
the upper right corners in Figure~\ref{fig:SMBHmassfunction}) are
overpredicted by a factor of $10^2-10^3$.  Note that these numbers
indicate the mass in SMBHs that avoided ejection due to kicks, and
remained in galactic centers (in some of the models, a significant
fraction of the BHs are ejected and form intergalactic BHs; there is
no obvious means to detect these~\cite{TH09}).  The mass density of
such nuclear SMBHs at $z\approx 0$ can be inferred from the observed
correlations between BH masses and host galaxy properties (such as the
masses or velocity dispersions of the host halos;
e.g.~\cite{Ferrarese02a}). The result is several$\times 10^5{\rm
M_\odot Mpc^{-3}}$; furthermore, the expectation is that most ($\gsim
90\%$) of this mass was accreted well after $z=6$
\cite{Shankarreview}.  Some strong feedback is therefore needed to
eliminate this significant overprediction. Possible candidates for
this are radiative feedback internal to halos, which maintains the
``$M-\sigma$ relation'' in ultra--high redshift, low-mass halos, or
the termination of PopIII star formation, at redshifts as high as
$z\sim 20$, due to Lyman Werner radiation \cite{HB06} or metal
pollution \cite{BFH09}.

Finally, it is worth emphasizing that the mass accretion rate
corresponding to the Eddington limit -- for the fiducial radiative
efficiency of $\epsilon\equiv L/\dot{m}c^2=0.1$ for converting mass to
radiation -- would need to be exceeded only by a factor of a $\sim$few
to make the growth from stellar seeds much easier.  Modestly exceeding
the Eddington rate is theoretically certainly plausible (see below):
density inhomogeneities can allow radiation to leak out of low density
regions while most of the accreting matter can be contained in high
density regions.  For example, magnetized radiation dominated
accretion disks are subject to a ``photon bubble'' instability that
nonlinearly appears to lead to strong density inhomogeneities
(e.g. \cite{Begelman02}). Nevertheless, observations have so far not
revealed systems that sustain super--Eddington accretion for extended
periods; it would then still have to be explained why the $z\approx6$
quasar BHs have this uniquely different behaviour.

\subsubsection{Accretion versus Mergers}   
\label{accretion_vs_mergers}

Mergers between halos can help build up the mass of individual black
holes (without significantly changing the total mass of the
population), provided that the central black holes in the halos
coalesce rapidly.  The mean accretion efficiency of $\sim 10$\%
inferred from comparing the local black hole mass density with the
integrated quasar light suggests that accretion dominates at least the
last e-folding of the black hole mass~\cite{YT02,Shankarreview}.
Mergers may, however, be significant earlier on~\cite{HCO04}.  In
addition, uncertainties in the {\it expected} radiative efficiency of
black hole accretion limit how accurately one can constrain the growth
of black hole mass by mergers. For example, if the typical efficiency
was $\approx 40$\%, as for a maximally rotating Kerr black hole, then
the Eddington-limited mass accretion rate would be decreased
correspondingly, and mergers could dominate black hole growth (on the
other hand, note that multiple mergers would have a tendency to cancel
the black hole spin; \cite{HB03}).  In order for mergers to contribute
significantly to the growth of individual black hole masses, stellar
seeds must be present in large numbers, in the most of the numerous
minihalos that form at $z\gsim 15$, down to small halo masses.

The balance between growth through BH mergers and growth through gas
accretion is indeed a key characteristic of any SMBH assembly
scenario.  For concreteness, consider possible merger histories for
the $z=5.82$ SDSS quasar SDSS 1044-0125 (\cite{HL01}, the following
arguments would be stronger for more luminous quasars at higher
redshift).  One can estimate the mass of the dark matter halo
harboring the quasar by its abundance. SDSS searched a comoving volume
of $\sim 1$ Gpc$^3$ to find each quasar. Assuming a duty cycle of a
few times $10^7$~years, one estimates that the dark matter halos
corresponding to this space density have masses of $10^{13}~\msun$
(using the halo mass function in ref.~\cite{Jenkins+01}, the original
Press-Schechter formula \cite{PS74} would give a similar answer).  A
$10^{13}~\msun$ halo at $z=6$ typically has only $\sim 10$ progenitors
with circular velocities of $v>50$~km~s$^{-1}$ (the other progenitors
being smaller).  This implies that mergers can only help build up the
black hole mass if seed black holes are present in progenitor halos
with much smaller masses.  A cutoff in the black hole mass function in
halos with circular velocities below $v=50$~km~s$^{-1}$ would be
justified if the cosmic ultraviolet background could suppress gas
infall into smaller halos
~\cite{Efstathiou92,TW96,NS97,KI00}. However, one-dimensional gas
collapse models with radiative feedback~\cite{Dijkstra+04} have shown
that this suppression is ineffective at redshifts beyond $z~\gsim 6$.
Thus, there is no obvious obstacle to forming seed black holes in
halos down to $v\sim 10$~km~s$^{-1}$ (below this threshold, atomic H
cooling becomes inefficient, and ${\rm H_2}$--photodissociation can be
a limitation).

In the illustrative models discussed in the previous section, which
successfully reproduce the abundance of the $\times 10^9{\rm M_\odot}$
SDSS quasar BHs, gas accretion accounts for the vast majority of the
growth (in the sense that if the seed BHs were simply added together
without any further accretion, the resulting total BH mass at $z=6$
would be reduced by many orders of magnitude; see Table 3 in
\cite{TH09}).  However, in versions of these models in which BH growth
is assumed to be self-regulating, accretion is much less important.
Such models essentially describe the most heavily merger-driven
scenarios possible, requiring accretion-driven growth of as little as
a factor of a few.  This is not surprising: placing a seed black hole
in each arbitrarily low mass progenitor halo, with the same black hole
mass to halo mass ratio as inferred for the SDSS quasars ($\mh/M_{\rm
halo}\sim 10^{-4}$), could account for the observed black hole masses
in quasars by $z=6$, even without {\em any} gas accretion
\cite{HCO04}.

A further important unsolved question is whether halo mergers
necessarily lead to black hole mergers at all (see, e.g., \cite{MM05}
for a review).  During a galaxy merger, the black holes sink via
dynamical friction to the center of the galaxy and form a tight black
hole binary in the nucleus.  In normal galaxies with a stellar
component in the nucleus, the black hole binary can continue to shrink
by ejecting low-angular momentum stars that pass close to the binary
(those in the ``loss cone'').  This process, however, clearly does not
operate in the earliest stages of structure formation, when there are
at most a few stars (if any) present in the merging mini-galaxies.
Even at the later states, this process is inefficient, at least in
spherical stellar systems, because the loss cone must be replenished
by two-body relaxation.  The black hole binary thus appears to stall
and cannot coalesce even during a Hubble time~\cite{BBR80}.

Several ideas for circumventing this difficulty have been proposed.
At later stages, in triaxial stellar populations, low-angular momentum
orbits are populated much more efficiently because the stellar orbits
can be chaotic; the resulting binary decay times are in many cases
significantly less than a Hubble time, even if only a few percent of
the stellar mass is on chaotic orbits (e.g.,
\cite{Yu02,MP04,Preto+11,Khan+11}).  In the earliest galaxies without
large stellar populations, the coalescence of BHs must be facilitated
by gas physics.  If circumbinary gas is present and forms a thin
accretion disk, then BH-disk interactions can drag the binary
together, in a manner similar to Type II migration in planetary
systems.  The main difference from the planet case is that at least
for nearly equal mass binaries, the secondary BH's mass will far
exceed the mass of the disk, which slows down the migration
(see~\cite{HKM09} for a comprehensive discussion, and
\cite{GR00,AN02,Escala+05} for examples of earlier work).  A binary BH
embedded in a spherical gas cloud is also facilitated by gaseous
torques~\cite{Escala+05}; this case is much less well explored, but is
likely to be more relevant to the earliest stages of the growth in the
stellar-seed models, in halos without stars and with potential wells
too shallow to support a thin disk.  Finally, if SMBHs are brought
together by successive halo mergers at a rate higher than the rate at
which they can coalesce, then one or more of the BHs can be ejected
out of the nucleus of the merger remnant by the slingshot
mechanism~\cite{SVA74}.  This could have implications for SMBH mass
build-up in principle; in practice, more recent work on the dynamics
of triple BHs indicate that ejections are relatively rare, and in the
majority of cases, at least two of the BHs
coalesce~\cite{HL07,Pau+10}.

\subsubsection{Super-Eddington Mass Accretion}   
\label{subsec:supereddingtonmdot}

If mass is supplied to a black hole at $\dot m \equiv \dot M/\dot
M_{Edd} \gg 1$, the photons are trapped in the inflowing gas because
the photon diffusion time out of the flow becomes longer than the time
it takes the gas to accrete into the black
hole~(e.g. \cite{Begelman78,BM82}).  The resulting accretion is thus
not via the usual thin disk~\cite{SS73}, but rather via a radiatively
inefficient flow (RIAF). The luminosity is still set by the Eddington
limit, but most of the gravitational binding energy released by the
accretion process is not radiated away (being trapped in the flow).

It is attractive to assume that the growth of SMBHs at high redshifts
proceeds via such an optically thick, photon trapped accretion flow
with $\dot m \gg 1$.  Indeed, it would be a remarkable coincidence if
the mass supply rate were precisely $\sim \dot M_{Edd}$ (required for
a thin accretion disk) during the entire growth of massive black
holes.  It is more likely that the mass supply rate is initially much
larger in the dense environments of high redshift galaxies ($\dot m
\gg 1$) and then slowly decreases with time as the galaxy is assembled
and the BH gains mass (e.g., \cite{SB92,CGM00}).  

Three-dimensional simulations for the cooling and collapse of gas into
the first minihalos find that ${\rm H_2}$ cooling reduces gas
temperatures to a few$\times 100$K and produces a quasi-static
contraction, with relatively low mass accretion rates of
$10^{-3}-10^{-2}~{\rm M_\odot~yr^{-1}}$ (e.g. \cite{ABN02}).  This
external mass-supply rate would still correspond to a super--Eddington
growth rate for a BH with a mass of $\lsim10^5{\rm M_\odot}$. However,
there is another limitation: within a Kelvin-Helmholtz time of $\sim
10^5$ years, only a few$\times100~{\rm M_\odot}$ of material is
accreted to the center (this is shown explicitly in
Figure~\ref{fig:Macc}). Much more mass than this is then unlikely to
be incorporated onto the central proto--star, before it settles to the
main sequence.  Radiative feedback from the proto--star (in the form
of ${\rm H_2}$ dissociation, Lyman $\alpha$ radiation pressure, and
ultimately, photoionization heating) on the infalling envelope was
found to limit the final mass of the star to $\approx 140~{\rm
M_\odot}$~\cite{TM04,MT08}.

On the other hand, in more massive halos {\em provided that $H_2$
cooling can be disabled throughout the entire time of the collapse},
the gas temperature is set by atomic H cooling and remains near
$10^4$K.  In a self-gravitating gas, the mass accretion rate is of
order $\sim c_s^3/G\propto T^{3/2}/G$ (e.g.~\cite{Shu77}).
Three-dimensional simulations have confirmed this scaling
(e.g.~\cite{ON07,SBH10}), and in halos with $T_{\rm vir}\sim 10^4$K,
have found mass accretion rates of $\sim 1 {\rm M_\odot~yr^{-1}}$
\cite{SBH10}.  As shown explicitly in Figure~\ref{fig:Macc}, with this
higher mass accretion rate, the mass that can be accumulated in the
nucleus within a Kelvin-Helmholtz time is increased to $10^5~{\rm
M_\odot}$.

Theoretical models for the accretion on much smaller spatial scales
(not resolved in the above simulations) imply that even if $\dot m \gg
1$, only a small fraction of the mass supplied to the black hole
actually reaches the horizon; most of it is driven away in an outflow
(see, e.g., simulations of RIAFs
\cite{SPB99,SP01,HB02,Igumenshchev+03,PB03}; and analytic models
\cite{BB99,BB04,QG00}).  The accretion rate onto a black hole thus
probably cannot exceed $\sim \dot M_{Edd}$ by a very large factor,
even if the mass supply rate from larger radii is large (see
\cite{SS73} for an early discussion of this point).

The above discussion focuses on whether highly super-Eddington
accretion is possible.  The question of whether the Eddington limit
for the luminosity can be exceeded by a modest factor of $\sim 10$ is
a bit more subtle.  Magnetized radiation dominated accretion disks are
subject to a ``photon bubble'' instability that nonlinearly appears to
lead to strong density inhomogeneities (see, in particular,
\cite{Arons92,Gammie98,Begelman01,BS01,Begelman02}).  Density
inhomogeneities allow super-Eddington fluxes from the accretion flow
because radiation leaks out of the low density regions while most of
the matter is contained in high density regions.
Ref.~\cite{Begelman02} estimates that the Eddington limit can
potentially be exceeded by a factor of $\sim 10-100$.  This would
allow much more rapid growth of black holes at high redshifts,
circumventing the above arguments that seed black holes at $z\sim 15$
are required.  Magneto-hydrodynamic (MHD) simulations of radiation
dominated accretion flows have confirmed the rapid growth of unstable
short-wavelength modes, with the development of large density
variations.  Inhomogeneities then allow the radiation to diffuse
outward five times more rapidly than in a disk in hydrostatic
equilibrium with no magnetic fields~\cite{Turner+05}.  Explicit models
for such slim, porous, accretion disks have been constructed
recently~\cite{DS11}.  In these models, when the external mass
accretion rate is 10-20 $\times \dot M_{Edd}$, despite the presence of
winds, a super-critical fraction, 2.6-3.8 $\dot M_{Edd}$, was indeed
found to reach the central SMBH.

\subsection{Growth by Rapid Direct Collapse}
\label{sec:directcollapse}

An appealing alternative idea is to produce, say, a $10^5~{\rm
M_\odot}$ SMBH ``directly'' -- i.e. much faster than this would take
under Eddington--limited accretion from a stellar seed.  This would
clearly be helpful to explain the high--redshift SMBHs.  In this
context, the crucial question is whether gas can accrete at a highly
super-Eddington rate onto a black hole, i.e., with $\dot M \gg \dot
M_{Edd}$, where $\dot M_{Edd} = 10 L_{Edd}/c^2 \approx 1.7
M_8~\msun$~yr$^{-1}$ is the accretion rate that would produce an
Eddington luminosity if accretion onto a black hole of mass $10^8
M_8~\msun$ proceeded with $10\%$ radiative efficiency. If so, this
could lead to rapid black hole growth at high redshifts.  Constraints
on BH seeds and their formation redshifts would therefore be much less
stringent.

\subsubsection{Rapid Collapse of Gas in $T_{\rm vir}\gsim 10^4$K Halos}   
\label{subsec:largehalos}

SMBHs may form directly by the collapse of gas clouds at high
redshifts, likely via an intermediate stage of a supermassive star or
disk.  The gas must not only shed angular momentum efficiently and
collapse rapidly, but must also then avoid fragmentation.  Whether
fragmentation of the gas cloud into stars can be avoided is
particularly questionable, in view of the large angular momentum
barrier that must be overcome to reach small scales in a galactic
nucleus (forming an SMBH through a dense stellar cluster is another
option, discussed in the next section).

The most promising locations for such rapid ``direct collapse'' are at
the centers of halos with $T_{\rm vir}\sim 10^4$K.  In the past
several years, many authors have sketched how gas may collapse
rapidly, without fragmentation, in these halos.  The essential idea is
that when contracting gas in a protogalactic nucleus becomes optically
thick and radiation pressure supported, it becomes less susceptible to
fragmentation and star formation.  It is, however, unlikely that
radiation pressure becomes important before angular momentum does,
implying that the gas forms a viscous accretion disk in the galactic
nucleus (fragmentation before the disk forms can also be avoided if
the forming fragments collide and ``coalesce'' before they can
separate into discrete dense clumps~\cite{KR83}).  On the other hand,
if self-gravitating, the resulting disk is strongly gravitationally
unstable and becomes prone to fragmentation and star formation (e.g.,
\cite{SB89,Goodman03}).  Whether this fragmentation can be avoided is
unclear. One possibility is to stabilize the disk by keeping its
temperature ``warm'' (i.e. $T\sim 10^4$K, close to the virial
temperature). This would fatten the disks (the scale height scales
with the ratio of gas and virial temperatures); this scenario may be
possible in a virtually metal--free, high redshift
halo~\cite{OH02,BL03,WTA08,RH09b}.

A suite of recent numerical simulations studied gas collapse in halos
with $T_{\rm vir}\sim 10^4$K~\cite{SBH10}. It was found found that the
gas in such halos, when collapsing in isolation, forms ${\rm H_2}$
efficiently, and (unfortunately) cools to temperatures of $T \sim
300$K.  Although no fragmentation was seen, the cold gas (well below
the virial temperature) is expected to ultimately fragment on smaller
scales that have not yet been resolved \cite{Turk+09}.  More
importantly, even if fragmentation was successfully avoided, there is
a problem: the cold gas was found to flow inward at relatively low
velocities, near the sound speed of $\sim 2-3~{\rm km~s^{-1}}$, with a
correspondingly low accretion rate of $\sim 0.01~{\rm
M_\odot~yr^{-1}}$.  \cite{SBH10} speculate that this is explained by a
series of weak shocks in the infalling gas, which prevent the gas from
accelerating to large Mach numbers (this is similar to the behavior
seen in three-dimensional simulations of the so-called ``cold mode''
of accretion in lower-redshift galaxies \cite{Keres+05}). Ultimately,
the slow infall velocities and cold temperatures produce conditions
nearly identical to those in the cores of lower-mass minihalos
(mentioned above); extensive ultra--high resolution simulations had
concluded that the gas then forms a single $\sim 100~{\rm M_\odot}$
star \cite{ABN02,BCL02,YOH08} or perhaps fragments even further into
several stars \cite{Turk+09,Stacy+10,Greif+11a,Prieto+11}, rather than
forming a supermassive star or BH.

There have been at least three different ideas on how to avoid ${\rm
H_2}$--cooling and to keep the gas warm.  One is for the gas to
``linger'' for a sufficiently long time at $10^4$K that it collapses
to a SMBH, even before ${\rm H_2}$ has a chance to reduce the
temperature (${\rm H_2}$ is kept dissociated by collisions before the
temperature falls below $\sim 4,000$K).  For a sufficiently high
space-- and column--density of neutral hydrogen, the absorption of
trapped Lyman $\alpha$ photons can be followed by collisional
de--excitation, rather than the resonant scattering of the Lyman
$\alpha$ photon, effectively trapping much of the cooling radiation.
This could prevent the gas temperature from falling below $\sim
8,000$K, and lead to such lingering and to SMBH formation -- analogous
to opacity--limited fragmentation in colder gas in the context of star
formation \cite{SS06,SSG10,Latif+11a}.

Another possibility is that, even in the presence of significant
cooling, angular momentum transport by gravitational instabilities,
spiral waves, bars, etc., can drive a fraction of the gas to yet
smaller scales in the galactic nucleus (e.g.~\cite{BS09}).
Ref.~\cite{EL95} argued that this was particularly likely to occur in
rare low angular momentum dark matter halos because the disk could
viscously evolve before star formation commenced.  A similar idea is
that a small fraction of the gas, with low specific angular momentum,
within the halo may collapse to the center without undergoing
fragmentation \cite{KBD04,LN06}.  It may help that even if most of the
gas is initially converted into stars, stellar winds and supernovae
will eject a significant amount of this gas back into the nucleus;
some of this gas can eventually collapse to smaller scales
\cite{BR78}.

\begin{figure}[t]
  \includegraphics[height=.6\textheight]{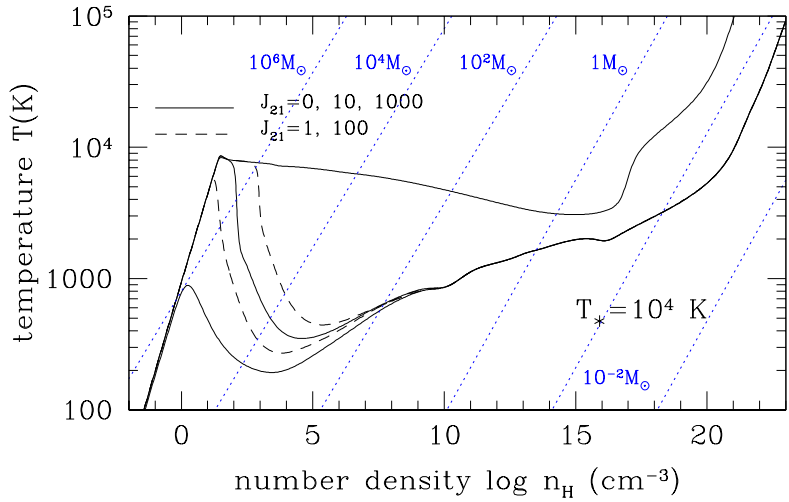}
  \caption{Temperature evolution of a metal-free cloud, irradiated by
    a strong UV flux.  The models solve for the chemical and thermal
    evolution, but assume a pre--imposed density evolution, based on
    the spherical collapse model.  Various cases are shown, with UV
    intensities at the Lyman limit of $J_{21}=0, 1, 10, 100$ and
    $10^{3}$, in the usual units of $10^{-21} {\rm
    erg~cm^{-2}~sr^{-1}~s^{-1}~Hz^{-1}}$ (solid and dashed curves; see
    the legend in the panel). Each blue dotted line corresponds to a
    different constant Jeans mass.  The gas is heated adiabatically
    until a density of $n\approx 10^0-10^2~{\rm cm^{-3}}$, at which
    ${\rm H_2}$--cooling becomes efficient and cools the gas to a few
    $\times 100$ K.  However, there exists a critical flux, with a
    value between $J_{21}=10^2$ and $10^3$, above which ${\rm
    H_2}$--cooling is disabled (adapted from ref.~\cite{OSH08}).}
\label{fig:jcritonezone}
\end{figure}

\begin{figure}[t]
  \includegraphics[height=.42\textheight]{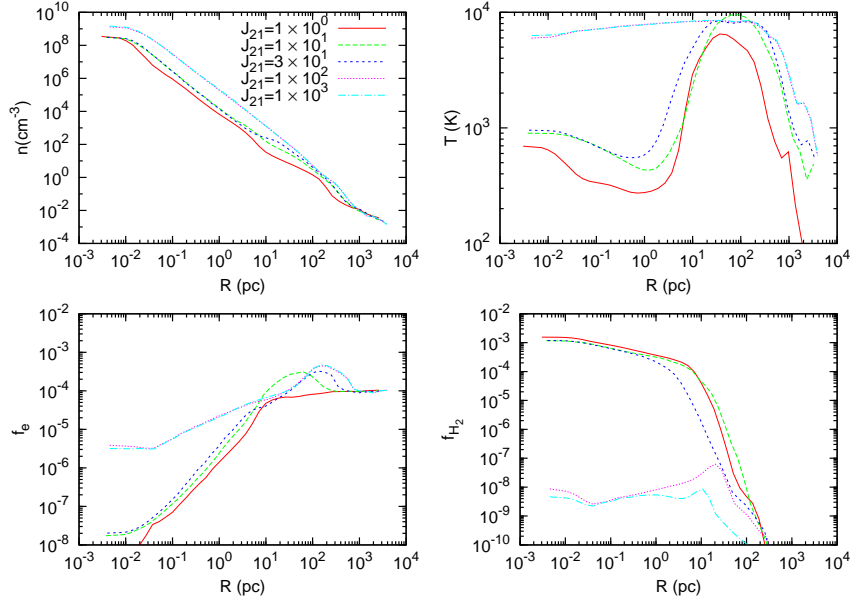}
  \caption{The results of adaptive mesh refinement (AMR) simulations
    of a primordial halo, with a total mass of a few $\times 10^7~{\rm
      M_\odot}$, collapsing at redshift $z\approx 10-15$, exposed to
    various UV background fluxes.  The four panels show snapshots of
    the spherically averaged profile of the particle number density,
    gas temperature, ${\rm e^{-}}$ fraction and ${\rm H_2}$ fraction,
    at the time of the collapse of the core, for several different
    values of the UV background intensity $J_{21}$, as labeled.  The
    existence of a critical flux, here with a value between
    $J_{21}=30$ and $10^2$, above which ${\rm H_2}$--cooling is
    disabled, is evident (adapted from ref.~\cite{SBH10}).}
\label{fig:jcritsim}
\end{figure}

Finally, ${\rm H_2}$--cooling may be disabled if the gas is exposed to
a sufficiently intense UV flux $J$, either directly
photo--dissociating ${\rm H_2}$ (in the Lyman--Werner bands near a
photon energy of $\sim 12$eV) or photo--dissociating the intermediary
${\rm H^-}$ (at photon energies $\gsim 0.76$eV).  Requiring the
photo-dissociation timescale, $t_{\rm diss}\propto J^{-1}$, to be
shorter than the ${\rm H_2}$--formation timescale, $t_{\rm
form}\propto \rho^{-1}$, generically yields a critical flux that
increases linearly with density, $J^{\rm crit} \propto \rho$.  In
low-mass minihalos, the critical flux is low, $J^{\rm crit}\approx
0.01-0.1$~\cite{HRL97,MBA01,MBH06,MBH09}\footnote{Here $J$ denotes the
specific intensity just below $13.6$eV, in the usual $J_{21}$ units of
$10^{-21} {\rm erg~cm^{-2}~sr^{-1}~s^{-1}~Hz^{-1}}$.}.  Since the gas
in halos with $T_{\rm vir}\gsim 10^4$K can cool via atomic Lyman
$\alpha$ radiation and loose pressure support, it inevitably collapses
further. As a result, in these halos, the critical flux is high,
$J^{\rm crit}\approx10^{2}-10^{5}$, depending on the assumed spectral
shape (ref.~\cite{SBH10}; see also refs. \cite{Omukai01,BL03} who
found similar, but somewhat higher values). The existence of this
critical flux is illustrated in Figure~\ref{fig:jcritonezone}, using a
one-zone model, in which the density evolution is approximated by
spherical collapse, and the gas is illuminated by a source with a
black-body spectrum (with a temperature of $T=10^4$K, characteristic
of a normal stellar population).  Figure~\ref{fig:jcritsim} shows the
radial structure of a $10^8{\rm M_\odot}$ halo, at the time of its
collapse, when illuminated at various intensities, taken from
three--dimensional simulations with the AMR code {\it enzo}.  These
profiles clearly show that when the UV flux exceeds a critical value,
the core of the halo is prevented from cooling to low temperatures.

\begin{figure}[t]
  \includegraphics[height=.4\textheight]{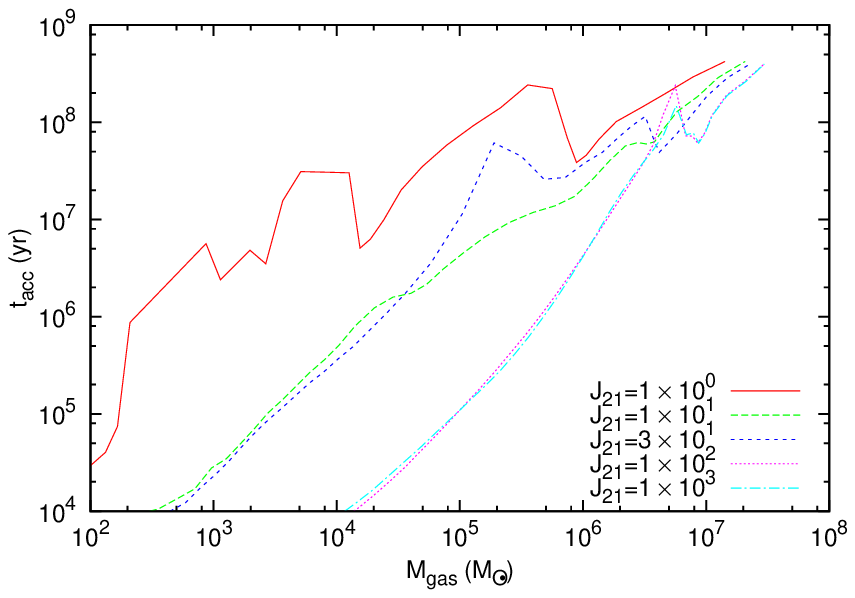} 
  \caption{The local accretion time--scale $t_{\rm acc}$ as a function
    of the enclosed gas mass $M_{\rm gas}$, in the same halo depicted
    in Figure~\ref{fig:jcritsim}, illuminated with different
    intensities, as labeled.  In the halos exposed to a supercritical
    flux ($J_{21}=10^2$ and $10^3$), the mass accretion rate,
    $\dot{M}\approx 1~{\rm M_\odot\,yr^{-1}}$, is nearly $10^3$ times
    higher than in halos whose gas cools via ${\rm H_2}$ ($J_{21}\lsim
    10$).  At the center of the brightly illuminated halos, $\approx
    10^5 {\rm M_\odot}$ of gas accumulates within a Kelvin-Helmholtz
    time of $\approx 10^5 {\rm M_\odot}$, possibly leading to the
    formation of a SMBH with a comparable mass (adapted from
    ref.~\cite{SBH10}).}
\label{fig:Macc}
\end{figure}

The 3D simulations also provide an estimate of the mass of the central
``object'' (star or SMBH) that ultimately forms at the core of the
halo, based on the measured profile of the mass accretion rate. This
is illustrated in Figure~\ref{fig:Macc}.  In particular, when the flux
exceeds the critical value, and the gas remains warm, the collapse is
significantly delayed.  However, when the gas ultimately does
collapse, it accretes toward the center at the sound speed
($c_s\approx 10$km/s), leading to a mass accretion rate of
$\dot{M}\approx 1{\rm M_\odot}\,{\rm yr^{-1}}$. This is much higher
than in the case of cold ($c_s\sim 1$ km/s) gas in halos with
efficient ${\rm H_2}$ cooling (as mentioned above, the simulations
find $\dot{M}\propto c_s^3$, as expected in self--gravitating gas).

Importantly, the critical flux is high -- likely significantly
exceeding the expected level of the cosmic UV background at high
redshifts.  Therefore, only a small subset of all $T_{\rm vir}\gsim
10^4$K halos, which have unusually close and bright neighbors, may see
a sufficiently high flux.  However, given the strong clustering of
early halos, there is a sufficient number of these close halo pairs to
account for the abundance of the rare $z=6$ quasars
\cite{Dijkstra+08}. A more significant challenge to this idea is that
in order to avoid fragmentation, the gas in these halos must also
remain essentially free of any metals and dust \cite{OSH08}.  This
requirement could be difficult to reconcile with the presence of a
nearby, luminous galaxies.

Finally, an important point to emphasize is that the collapsing gas is
optically thick, and the critical flux $J_{\rm crit}$ depends
crucially on the details of self-shielding of the Lyman-Werner lines
of ${\rm H_2}$.  Since following radiative transfer in many dozens of
lines is computationally expensive, existing works have employed
various simplifying approximations.  The simplest (and by far most
commonly used) approach is to combine a simple power-law fitting
formula, $f_{\rm shield}=(N_{\rm H_2}/10^{14}{\rm cm^{-2}})^{-3/4}$,
for the ${\rm H_2}$ self-shielding factor~\cite{DB96} with an estimate
for an effective ${\rm H_2}$ column density $N_{\rm H_2}$ (most often
equated with the product of the local density and Jeans length).
These assumptions have recently been scrutinized in ref~\cite{Jemma2},
which solved radiative line transfer exactly, using post-processing of
three-dimensional simulations.  This showed, rather promisingly, that
when self-shielding is treated more accurately, $J_{\rm crit}$ is
reduced by about an order of magnitude.  Interestingly, this reduction
comes from a product of three very different sources (each of which
individually reduces the shielding by a factor of $\sim$two): (1) a
numerical inaccuracy of the power-law $f_{\rm shield}$ formula, (2)
the inapplicability of this fitting formula at the relevant,
relatively high temperatures ($\gsim 10^3$K), where excited rotational
levels of ${\rm H_2}$ are populated, and (3) the Jeans length yielding
an overestimate of the effective average (over different sightlines)
column density.  The order-of-magnitude reduction in $J$ is especially
important, since the probability distribution of $J$, sampled by halos
at $z\gsim 10$, is very steep near $J\sim 10^4$ (see
\cite{Dijkstra+08}).  With the original high $J_{\rm crit}$ value, it
has been shown \cite{Dijkstra+08} that only one in $\approx10^6$ halos
-- only those with an unusually bright and close neighbour -- will see
a sufficiently high flux.  The reduction of the $J_{\rm crit,21}$
value will significantly increase the number of candidates for objects
that can avoid ${\rm H_2}$--cooling and fragmentation, and makes this
scenario much more viable.

\subsubsection{The Ultimate Fate of the Gas}
\label{zoltansubsec:collapse}

Although the detailed evolutionary pathways are still not understood,
a possible outcome of the above scenarios is the continued collapse of
some gas to smaller scales in the galactic nucleus.  As the gas flows
in, it becomes optically thick, and the photon diffusion time
eventually exceeds the inflow time.  Radiation pressure dominates for
sufficiently massive objects so that the adiabatic index is $\Gamma
\approx 4/3$.  Radiation pressure may temporarily balance gravity,
forming a supermassive star or disk (SMS; e.g., \cite{HF63,Wagoner69};
see, e.g., \cite{ST83} for a review and additional references to
earlier work).  The SMS will radiate at the Eddington limit and
continue contracting.  When the SMS is sufficiently compact ($GM/Rc^2
\approx 10^{-4} M_8^{-1/2}$ for non-rotating stars), general
relativistic corrections to the gravitational potential become
important, and the star becomes dynamically unstable because its
effective polytropic index is $\lsim 4/3$.  For masses $\lsim
10^5~\msun$, thermonuclear reactions halt the collapse and generate an
explosion (e.g., \cite{FWW86}), but more massive objects appear to
collapse directly to a SMBH (see \cite{Shapiro04} for a review; and,
e.g., \cite{SS02,Saijo+02} for recent simulations).

If the mass accretion rate is high ($\sim 1~{\rm M_\odot yr^{-1}}$),
the outer layers of the SMS do not have time to thermally relax, and a
high-pressure core-envelope structure may develop, dubbed a
``quasi-star'' \cite{BVR06,BRA08,Begelman10}. The envelope initially
contains most of the mass, and the central BH embedded in the envelope
can grow from this envelope; the key feature of this configuration is
that the accretion is limited by the Eddington limit for the entire
envelope, rather than just the BH.  Interestingly, the mass accretion
rate required for this model is comparable to that seen in
three-dimensional simulations of $T_{\rm vir}\gsim 10^4$K halos, {\em
provided again that the gas in these halos can avoid $H_2$
cooling}~\cite{SBH10}.

Finally, a possibility for the gas is to ultimately fragment into
stars, but not before it reaches very high densities.  If the gas is
metal--enriched, this scenario may be most likely, and will result in
the formation of a dense and compact stellar cluster, which naturally
evolves to form a SMBH~\cite{OSH08}.  We turn to this idea in the next
section.

\subsection{The Formation of Black Holes in Stellar Clusters}
\label{zoltansubsec:clusters}

The negative heat capacity of self-gravitating stellar systems makes
them vulnerable to gravitational collapse in which the core of the
cluster collapses on a timescale $t_{cc}$ comparable to the two-body
relaxation time of the cluster \cite{BT87}.  If core collapse proceeds
unimpeded, the resulting high stellar densities can lead naturally to
the runaway collisional growth of a single massive object which may
evolve to form a black hole (as in the discussion of SMSs above).
This process provides an additional route for the direct formation of
SMBHs at high redshifts (or, more likely, intermediate mass seeds).

Early work suggested that the fate of stellar clusters depends
sensitively on the number of stars in the cluster.  \cite{Lee87} and
\cite{QS90} found that very dense massive star clusters ($N~\gsim
10^6-10^7$ stars) were required to have successful core collapse and
runaway growth of a single massive object.  In less massive clusters,
core collapse was halted by binary heating, in which the cluster gains
energy at the expense of binaries via three-body interactions
\cite{Heggie75,Hut+92}.  Successful core collapse also requires that
$t_{cc}$ is shorter than the timescale for the most massive stars to
evolve off the main sequence (\cite{RFG04}; this requirement implies
compact clusters $\lsim 1$~pc in size). Otherwise, mass loss from
evolved stars and supernovae prevents the core from collapsing (in
much the same way as binary star systems can become unbound by
supernovae). In principle, massive stars could evolve into
stellar-mass BHs and form a dense cluster of stellar--mass BHs. In the
context of high-redshift halo formation, this is naturally
expected~\cite{MR01}, given that the first stars in the first
minihalos are believed to be massive.  A dense cluster of
stellar--mass BHs can, in principle, grow into a more massive IMBH by
coalescence due to gravitational radiation; however, this process is
effective only in large stellar systems that are found in present-day
galactic nuclei~\cite{OLeary+06}.

If the required number of stars was indeed as large as $N~\gsim
10^6-10^7$, this would be bad news for early SMBH formation: in the
cosmological hierarchy, $\gsim 10^8~{\rm M_\odot}$ halos -- the
smallest that could plausibly harbor such star clusters -- are very
rare ($\gsim 2.5\sigma$ fluctuations) at $z\gsim 10$.  However, recent
work has revived earlier ideas that stellar clusters are subject to a
``mass segregation instability'' that makes even the relatively less
massive clusters prone to forming black holes
(\cite{Spitzer69,Vishniac78,BR78}).  Because massive stars in a
cluster sink by dynamical friction toward the center (mass
segregation), they invariably dominate the dynamics of the cluster
core and can undergo core collapse on a timescale much shorter than
that of the cluster as a whole (and on a timescale shorter than their
main sequence lifetime).  \cite{PM02} showed with N-body simulations
that the resulting core collapse likely leads to runaway merger and
formation of a single black hole, and \cite{GFR04} reached a similar
conclusion for much larger $N \sim 10^7$ using Monte Carlo
simulations.  Including an explicit treatment of stellar collisions,
the most recent Monte-Carlo simulations find that the central massive
object that forms in core has a mass of $\sim 10^{-3}$ of the whole
cluster \cite{Goswami+11}.

In the context of high-redshift BH formation, \cite{OSH08} considered
the cooling properties of gas in $T_{\rm vir}\gsim 10^4$K halos.  It
is expected that the majority of such halos, when they are assembled,
had already undergone some amount of star--formation. In this case,
their gas would have at least a trace amount of metals.  \cite{OSH08}
considered the case when such mildly polluted halos are exposed to a
large UV flux, which dissociates ${\rm H_2}$, and initially prevents
cooling.  This allows the gas to contract to very high densities,
without fragmenting initially.  By following the thermal and chemical
evolution of such low--metallicity gas, exposed to extremely strong UV
radiation, \cite{OSH08} found, however, that eventually, gas
fragmentation is inevitable above a critical metallicity, whose value
is between $Z_{\rm cr} \approx 3 \times 10^{-4} Z_{\odot}$ (in the
absence of dust) and as low as $Z_{\rm cr} \approx 5 \times 10^{-6}
Z_{\odot}$ (with a dust-to-gas mass ratio of about $0.01
Z/Z_{\odot}$).  When the metallicity exceeds these critical values, an
ultra--dense cluster (the density at the time of fragmentation is
$n\gsim 10^{10}~{\rm cm^{-3}}$) of low--mass stars may form at the
halo nucleus (\cite{Clark+08} and \cite{DV09} argued for similar
scenarios, in the central regions of a protogalactic disk).
Relatively massive stars in such a cluster can then rapidly coalesce
into a single more massive object, which may produce an
intermediate--mass BH remnant with a mass up to $M\lsim 10^3~{\rm
M_\odot}$.

The above processes provide a promising channel for the formation of
IMBH seeds, which can grow via mergers and/or accretion to form SMBHs.
For example, \cite{VHM03} and \cite{Islam+03} have incorporated such
early black hole seeds into Monte Carlo simulations of the black hole
merger histories. With reasonable prescriptions for the merging and
accretion of black holes inside dark halos, these models can account
for the observed evolution of the quasar luminosity functions at $z<5$
and can serve for physically motivated extrapolations to high
redshifts to describe the first AGN.

It should be noted that there exist some observational evidence for
IMBHs in the local universe.  In particular, the presence of IMBHs
with masses of order $\sim 10^4~{\rm M_\odot}$ are inferred from
stellar kinematics in the globular clusters G1 (\cite{Gebhardt+02};
note that the velocity dispersion profile itself does not require a
BH~\cite{Baumgardt+03} and the evidence comes from higher-order
moments of the velocity distribution \cite{Gebhardt+05} instead),
$\omega$ Cen~\cite{Noyola+08} and in M15 (\cite{vanderMarel+02},
although this object can also be modeled without an IMBH;
\cite{vanderMarel04}).  Ultra-luminous X-ray sources in nearby
galaxies (e.g., \cite{CM99,Kaaret+01,MC04,Farrell+09}) have also been
interpreted as accreting IMBHs.  While there are viable non-IMBH
interpretations of these sources (e.g., \cite{King+01,Begelman02}),
the X-ray spectrum of one such source during the peak of an outburst
implies the presence of a $\gsim 2000~{\rm M_\odot}$ IMBH (assuming
that the luminosity is limited to $0.3\times L_{\rm Edd}$ as in
stellar-mass BH X-ray binaries in their hard state; \cite{Kaaret+09}).

\subsection{Alternative Models}
\label{subsec:others}

Since both of the ``standard'' scenarios discussed above require some
optimistic assumptions, it is interesting to consider some more exotic
possibilities.

\begin{figure}
  \includegraphics[height=.6\textheight]{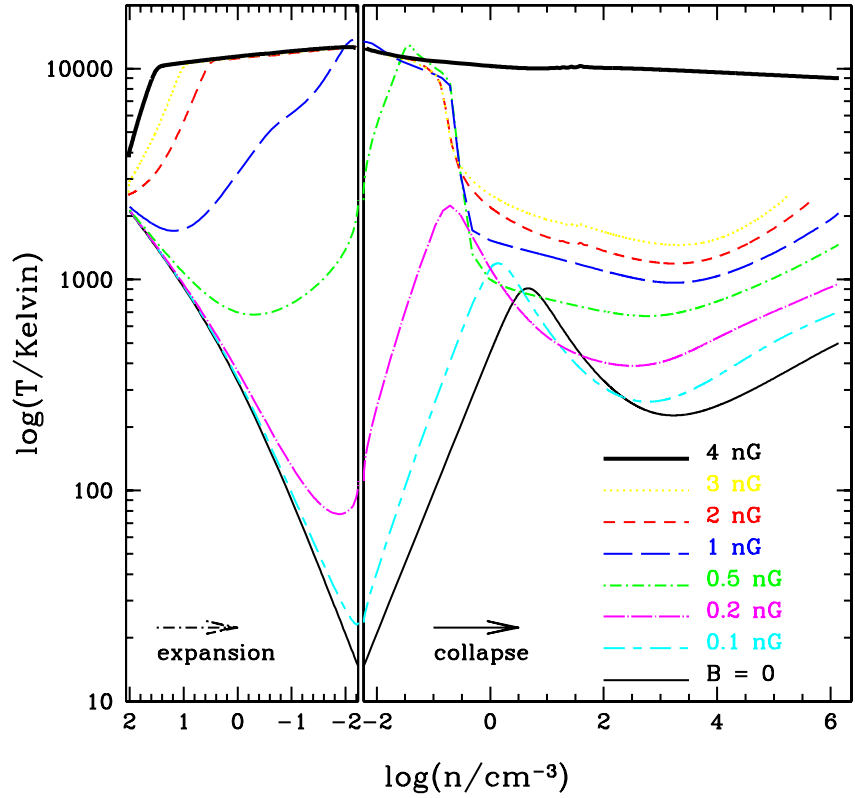}  
  \caption{The temperature evolution of a patch of the intergalactic
    medium is shown as it initially expands and then turns around and
    collapses to high density.  The different curves correspond to
    different values of the assumed primordial magnetic field, as
    labeled.  The gas evolves from the left to the right on this
    figure.  The left panel shows the expanding phase, starting from
    an initial density of $\approx 100~{\rm cm^{-3}}$ (corresponding
    to the mean density at redshift $z\simeq 800$) and ending at the
    turnaround just below $n=10^{-2}~{\rm cm^{-3}}$. The right panel
    follows the subsequent temperature evolution in the collapsing
    phase.  The figure shows the existence of a critical magnetic
    field, with a value between $B=3$ and $4$ nG, above which ${\rm
    H_2}$--cooling is disabled, and the gas temperature always remains
    near $10^4$K (adapted from ref.~\cite{SHP10}). }
\label{fig:Bfield}
\end{figure}

It is commonly believed that the magnetic fields permeating galaxies
such as the Milky Way arose by the amplification of a much weaker
large--scale seed field.  Weak primordial magnetic fields, with
strengths of up to $\sim$ 1nG, can be produced in phase transitions in
the early universe, during inflation, or during the electroweak or QCD
phase transitions.  It has recently been shown that such a primordial
magnetic field could produce a variant of the ``direct collapse''
scenario \cite{SHP10}. In particular, if the field is tangled, then
ambipolar diffusion will provide an efficient new mechanism to heat
the gas as it collapses in protogalactic halos. If the field has a
strength above $\mid B\mid\gsim 3$ (comoving) nG, the collapsing gas
is kept warm ($T\sim 10^4$ K) until it reaches the critical density
$n_{\rm crit}\approx10^3 {\rm cm^{-3}}$ at which the roto--vibrational
states of ${\rm H_2}$ approach local thermodynamic equilibrium.  This
is illustrated explicitly by the thermal evolution of fluid elements
shown in Figure~\ref{fig:Bfield}.  ${\rm H_2}$--cooling then remains
inefficient, and the gas temperature stays near $\sim 10^4$K, even as
it continues to collapse to higher densities. The critical magnetic
field strength required to permanently suppress ${\rm H_2}$--cooling
is somewhat higher than upper limit of $\sim 2$nG from the cosmic
microwave background (CMB). However, it can be realized in the rare
$\gsim(2-3)\sigma$ regions of the spatially fluctuating $B$--field;
these regions contain a sufficient number of halos to account for the
$z\approx6$ quasar BHs
\footnote{Because of the high magnetic Jeans mass, the magnetic
pressure has significant dynamical effects, and can prevent gas
collapse in halos with masses up to $M\gsim {\rm few}\times
10^{10}{\rm M_\odot}$. These are $\sim$100 times more massive than the
DM halos in the ``usual'' direct collapse models discussed in
\S~\ref{sec:directcollapse} above.}

Another ``exotic'' idea is that the first PopIII stars may be powered
by heating by dark matter annihilation, rather than by nuclear fusion
\cite{Spolyar+08}.  Weakly interacting massive particles (WIMPs), can
be such a heat source, as long as they reach sufficiently high density
inside the first stars, and if the annihilation products are trapped
inside the star.  Several authors have studied the impact of this
additional heating mechanism on the structure and evolution of such
``dark stars'' \cite{Spolyar+09,Iocco+08,Yoon+08,Taoso+08,Umeda+09,
Spolyar+09,Freese+10,Ripamonti+10}.  In particular, these stars can
live much longer than ``normal'' PopIII stars, and because their
radiation is soft, they can continue to accrete gas, as long as the
dark matter heating persists, and grow to masses of up to $\sim
10^5{\rm M_\odot}$. In fact, one of the challenges in these models is
to explain why and how the growth of the star stops
\cite{Umeda+09,Freese+10}.  An interesting prediction is that these
stars are bright, and should be detectable directly by {\it
JWST}~\cite{Freese+10}.

\section{Observational Considerations}
\label{sec:observations}

In this section, we first review several recent observations and their
implications for the formation of black holes at high redshifts. We
then then speculate on how future observations may probe the assembly
of high-$z$ SMBHs.

\subsection{Surveys for High Redshift Quasars}
\label{zoltansubsec:sdss}

The majority of the $\sim 40$ quasars known at $z\sim 6$ to date have
been discovered in the SDSS.  This is perhaps somewhat surprising,
since the SDSS is a relatively shallow survey (with a magnitude limit
of $i\sim 22$) capable of detecting only the rarest bright quasars at
redshifts as high as $z\sim 6$.  Nevertheless, the large solid angle
searched for high redshift quasars ($\sim 8,000$ square degrees) has
yielded many such objects \cite{Fanreview06}.  Somewhat deeper surveys
covering smaller areas (a few $\sim 100$ square degrees), such as the
SDSS Deep Stripe \cite{Jiang+09}, and the CFHQS~\cite{Willott+10a} and
UKIDSS~\cite{Lawrence+07} surveys, have yielded many additional
detections.  The most important properties (for our purposes) of these
sources are that they are probably powered by SMBHs as large as a few
$\times~10^9~\msun$ and overall, they appear to be indistinguishable
from bright quasars at moderate ($z\sim 2-3$) redshifts, with similar
spectra and inferred metallicities (e.g. \cite{Jiang+07}. In addition,
a large reservoir of molecular gas is already present, even in the
most distant sources \cite{Walter+03,Wang+10,Wang+11}.  

Despite the overall similarities, there are some tentative
distinctions between these $z\gsim 6$ quasars and their $z\sim 2-3$
counterparts. First, there is evidence for increasing Eddington ratios
toward high redshift (see below).  Second, quasar clustering has been
found to strongly increase with redshift \cite{Shen+07}.  The observed
clustering strength can be used to infer to quasar lifetimes
\cite{HH01,MW01}, and implies that the duty cycle of bright quasar
activity increases significantly toward high redshifts, to near unity
by $z\approx 6$~\cite{Shankar+10}.  Finally, two of the $\sim 6$
quasars have no detectable emission from hot dust \cite{Jiang+10}.
There are no known examples for such apparently hot-dust-free quasars
at low redshift; this result therefore suggests that at least these
two quasar BHs may have been caught at a young age of the evolution of
their host galaxies (i.e., there was insufficient time for a dusty
torus to form in the nucleus).

With these exceptions, the high-$z$ SMBHs and their surroundings
appear as ``fully developed'' as their lower redshift counterparts,
despite the young age ($\lsim 10^9$~years) of the universe at $z\gsim
6$.  These rare quasars are likely harbored by massive ($\sim
10^{12-13}~\msun$) dark matter halos that form out of $4-5\sigma$
peaks of the fluctuating primordial density field. The large halo mass
follows directly from the space density of these sources (\cite{HL01};
another method to confirm the large halo masses is to study the
expected Lyman $\alpha$ absorption signatures of cosmological gas
infall onto such massive halos, as proposed by \cite{BL03}).  Indeed,
the environment and dynamical history of an individual massive dark
matter halo at $z\sim 6$ and $z\sim 3$ can be similar; it is their
abundance that evolves strongly with cosmic epoch. This is broadly
consistent with the observations: the bright $z\sim 6$ quasars look
similar to their $z\sim 3$ counterparts, but their abundance is much
reduced (by a factor of $\sim 40$).

The fact that these quasars are so rare has important implications.
First, they are likely to be the ``tip of the iceberg'' and
accompanied by much more numerous populations of fainter quasars at
$z\gsim 6$.  The slope of the luminosity function is expected to be
very steep at $i\sim 22$, and so pushing the magnitude limits further
in future surveys should prove rewarding. The most direct constraints
on this slope are from combining the CFHQS sample with the more
luminous SDSS sample (yielding a total of 40 quasars between redshifts
$5.74 < z < 6.42$; \cite{Willott+10a}), and from gravitational lensing
\cite{CHS02,WL02a,WL02b,Richards+04}. Combining source counts and
lensing yields the strongest limit of $-d\log\Phi/d\log L\lsim
3$~\cite{Wyithe04}.  Second, the steep slope of the dark halo mass
function implies that the masses of the host halos can be ``measured''
from the abundance quite accurately (see discussion in
\S~\ref{zoltansubsec:fossils}).  Conversely, since small changes in
the assumed host halo mass results in large changes in the predicted
abundance, large uncertainties will remain in other model parameters.
In this sense, fainter, but more numerous quasars (or lack thereof)
can have more constraining power for models that relate quasars to
dark halos.

The most striking feature of the SDSS quasars, however, is the large
black hole mass already present at $z\sim 6$.  In the rest of this
section, we critically assess whether the inferred large black hole
masses are robust.

The masses of the black holes powering the SDSS quasars are inferred
by assuming that (1) they shine at the Eddington luminosity with a
bolometric correction identical to that of lower redshift quasars
(this is justified by their similar spectra), and (2) they are neither
beamed nor gravitationally lensed (both of these effects would make
the quasars appear brighter and allow lower BH masses).  These
assumptions lead to black hole masses $\mh\approx(2-6)\times 10^9~{\rm
M_\odot}$ for the $z>6$ quasars known to date.  These are reasonable
assumptions, which have some empirical justification.

The hypothesis that the quasars are strongly beamed can be ruled out
based on their line/continuum ratio. If the quasar's emission was
beamed into a solid angle covering a fraction, $f$, of $4\pi$, it would
only excite emission lines within this cone, reducing the apparent
line/continuum ratio by a factor $f$. However, the SDSS quasars have
strong lines. \cite{HC02} found that the line/continuum ratio of the
$z=6.28$ quasar SDSS 1030+0524 is about twice that of the median value
in the SDSS sample at $z>2.25$ \cite{VandenBerk+01}. This argument,
applied to the Mg~II line of the $z=6.41$ quasar SDSS J1148+5251
\cite{WMJ03} yields a similar conclusion.

Another important uncertainty regarding the inferred black hole masses
is whether the SDSS quasars may be strongly magnified by gravitational
lensing.  The optical depth to strong lensing along a random line of
sight to $z\sim 6$ is small ($\sim 10^{-3}$; e.g.,
\cite{Kochanek98,BL00}).  Nevertheless, magnification bias can
significantly boost the probability of strong lensing.  If the
intrinsic (unlensed) luminosity function at $z\sim 6$ is steep and/or
extends to faint magnitudes, then the probability of strong lensing
for the SDSS quasars could be of order unity
\cite{CHS02,WL02a,WL02b}). The overwhelming majority (more than 90\%)
of strong lensing events would be expected to show up as multiple
images with separations at least as large as $0.3''$. It is difficult
to produce strong magnification without such multiple images, even in
non-standard lensing models (allowing ellipticity and/or external
shear \cite{KKH05}). However, deep optical observations (e.g. with the
{\em Hubble Space Telescope}) of the highest redshift quasars show no
signs of multiple images for any of the $z~\gsim 6$ sources down to an
angle of $0.3''$ \cite{Richards+04,Willott+05}.\footnote{The
highest-redshift known lensed quasar is at $z=4.8$ and was discovered
serendipitously in SDSS, initially flagged as a galaxy due to the
strong blending of one of the quasar images with a bright galaxy
\cite{McGreer+10}.}

Finally, whether or not the SDSS quasars are shining at the Eddington
limit is difficult to decide empirically.  Several authors
\cite{WMJ03,Vestergaard04,Jiang+07,Kurk+07,Vestergaard+08,Vestergaard+09,Kurk+09,Willott+10b}
have estimated Eddington ratios in samples of high redshift quasars,
using observed correlations between the size of the broad line region
and the luminosity of the quasar (the correlation is calibrated using
reverberation mapping of lower redshift objects;
e.g. \cite{Kaspi+00,Vestergaard02,Kollmeier+06}). Values range from
$\approx 0.1$ to $\gsim 1$; in particular, $L/L_{Edd}$ is typically
found to increase with redshift, and approaches unity for the $z~\gsim 6$ quasars.

Inferences about Eddington ratios at high redshifts can also be made
by utilizing models of the quasar population as a whole.  Such models
typically assume the Eddington luminosity at higher redshifts, where
fuel is thought to be readily available \cite{SB92,HR93}. Numerous
semi-analytic models for the quasar population (see, e.g.,
ref.~\cite{Shankarreview} and references therein) have found that
Eddington ratios of order unity during most of the growth of the black
hole mass also yield a total remnant SMBH space density at $z=0$ that
is consistent with observations.  Ref.~\cite{CO01}, and, more
recently, \cite{Milos+09} have self-consistently modeled accretion and
radiative feedback onto an individual quasar BH, and found that
(provided fuel is available) the luminosity is near the Eddington
value during the phases when the quasar is on.  Despite these
arguments, one cannot directly rule out the possibility that the SDSS
quasars shine at super-Eddington luminosities (theoretically, this is
possible, as discussed above). We emphasize that if this were true and
the masses were lower than $10^9~\msun$, then the SDSS quasars would
have to be luminous for only a short time: maintaining the observed
luminosities for $\gsim 10^7$ years with a radiative efficiency of
$\epsilon\equiv L/\dot m c^2=0.1$ would bring the black hole masses up
to values of $10^9~\msun$ anyway.

\subsection{Local Black Holes as Fossils}
\label{zoltansubsec:fossils}

As mentioned above, SMBHs appear ubiquitous in local galaxies, with
their masses correlating with the global properties of their host
spheroids. Several groups have noted the broad natural implication
that the formation of the SMBHs and their host spheroids must be
tightly linked (see, e.g., \cite{Shankarreview}).  Various independent
lines of evidence suggest that spheroids are assembled at high
redshifts ($z\sim 2$; see \cite{CB03} for the age determinations from
the Sloan sample and references to older work), which would be
consistent with most of the SMBH mass being accreted around this
redshift (coinciding with the peak of the activity of luminous
quasars). Indeed, starting from the age distribution of local
early-type Galaxies, one can reconstruct the cosmic evolution of the
quasar luminosity function to within observational errors, using the
most naive set of assumptions (namely that the formation of stars and
the assembly of the nuclear SMBHs track one another, with the SMBH
radiating at a constant $f_{\rm Edd}\sim 0.3$, and that the
$\mh-\sigma$ correlation does not evolve with
redshift~\cite{HJB07,SBH09}).

This then has the unwelcome (but unsurprising) implication that the
local SMBHs may contain little direct evidence of the formation of
their seeds at $z>6$.  Indeed, it seems most plausible that the
observed tight correlations, such as between $\mh$ and $\sigma$, are
established by a feedback process which operates when most of the
black hole mass is assembled.  However, an upside of this hypothesis
is that---with the identification of a specific feedback
mechanism---physically motivated extrapolations can be made toward
high redshifts.  Also, while relative massive local SMBHs ($\sim
10^9~{\rm M_\odot}$) have undergone many mergers, those with the
lowest masses ($\sim 10^6~{\rm M_\odot}$) are more likely to have
avoided mergers.  Therefore, the low-mass end of the $\mh-\sigma$
relation could be a probe of high-$z$ SMBH formation models.  To be
more explicit: if only a small fraction of high-$z$ halos are seeded
with BHs, massive galaxies will have undergone many mergers, and will
have a nuclear SMBH at present (i.e. $f_{\rm seed}\approx 1$).  On the
other hand, many low--mass galaxies may still have no BHs and thus the
dwarf-galaxy population can have $f_{\rm seed}\ll 1$
\cite{MHN01}. Likewise, the direct-collapse models produce SMBHs whose
masses are initially well above the $\mh-\sigma$ relation, which can
be a diagnostic of such models \cite{TH09}. Such an 'upward curvature
in the low-mass end of the relation could indeed be preserved all the
way down to $z=0$ \cite{VN09}.

More generally, whether the local $\mh-\sigma$ relation holds at
higher redshifts, both in normalization and in slope (as discussed by
several authors), and also in range (which has received less
attention, but see \cite{Netzer03}), are interesting observational
questions.  The highest redshift SDSS quasars do appear to satisfy the
$\mh-\sigma$ relation of the local SMBHs, at least approximately. If
$\mh$ is estimated assuming the Eddington luminosity, and $\sigma$
from the CO line-width, then, at a given $\sigma$, high-$z$ quasars
have BH masses a factor of $\sim 4$ larger than local galaxies,
although this could still be partly caused by selection (with high-$z$
quasars preferentially viewed face-on)~\cite{Ho07} and by the
observations probing a smaller (inner) fraction of the DM halo.  We
also note that if $\sigma$ is estimated from the circular velocity of
the host dark matter halos with the right space density (e.g.,
\cite{HL01}), then the SDSS quasars are within the scatter of the
$\mh-\sigma$ relations of \cite{Gebhardt+00} and also of
\cite{Ferrarese02a}.  As explained in \S~\ref{zoltansubsec:sdss}, the
halo mass inference is reasonable.  The determination of the halo mass
and circular velocity from the observed abundance of quasars is also
more robust than it may at first appear. This is because, despite the
dependence on the poorly known duty cycle, the halo mass function is
exponentially steep for the massive $M\sim 10^{13}~\msun$ halos at
$z\sim 6$; therefore, the dependence of the inferred halo mass on the
duty cycle (and other uncertainties in the estimated halo abundance)
is only logarithmic.  The weakest link in the argument is associating
the spheroid velocity dispersion with the circular velocity of the
dark matter halo.  There is evidence~\cite{Ferrarese02a} of a
correlation between $\mh$ and $\sigma$, with the velocity dispersion
measured in the dark matter dominated region of SMBH host galaxies;
this establishes a direct link to the dark halo and puts the above
argument on somewhat firmer ground (although there are still large
errors in the inferred correlation, depending on the halo profile one
adopts to convert the measured circular velocity to total halo mass).

The (tentative) evidence that high-redshift AGN do not strongly
deviate from the local $\mh-\sigma$ relation further supports the idea
that the formation of SMBHs and their host galaxies must be tightly
coupled by cosmology-independent physical processes (since the SDSS
quasars are the rare peaks that have already formed at $z\sim 6$
instead of at $z\sim 2$).  Besides the slope and normalization of the
$\mh-\sigma$ relation, the {\it range} (of masses and velocity
dispersions) over which observed galaxies satisfy this relation has to
match up between low and high redshifts. In particular, the largest
black holes observed at high redshifts have inferred masses
approaching $\mh \sim 10^{10}~\msun$.  These should also exist at low
redshifts, but have not yet been discovered. In the SDSS, the galaxy
with the velocity dispersion record has $\sigma=444~{\rm
km~s^{-1}}$\cite{Salviander+08}), whereas a naive application of the
local $\mh-\sigma$ relation would predict the presence of $\sigma>700
~{\rm km~s^{-1}}$ galaxies \cite{Netzer03}.  This puzzle is alleviated
somewhat by the scatter in the relation; it is likely fully resolved
by the realization that the $\mh-\sigma$ relation has 'curvature',
with BHs in the largest galaxies preferentially more massive than the
power-law $\mh-\sigma$ relation would predict~\cite{Lauer+07}.

There have been several suggestions in the literature for the nature
of the dynamical coupling between the formation of the black hole and
its spheroid host.  The most promising is radiative or mechanical
feedback from the SMBH on the gas supply in the bulge.  The essential
idea (going back to \cite{SR98}) is that when the black hole in the
center of the galaxy grows too large, its outflows and radiation
unbind the gas in the bulge or in the disk, quenching further black
hole growth via accretion and further star formation.  Competition
with star formation for the gas supply may also play a
role~\cite{DiMatteo+03,LHM07}.  Note that these mechanisms can readily
work at any redshift.

There are several alternative possibilities for the origin of the
$\mh-\sigma$ relation, which include: (1) filling the dark matter loss
cone \cite{Ostriker00}.  In this model, the growth of the SMBH occurs
first through the accretion of collisional dark matter particles, and
subsequently through the scattering of these particles into orbits
that are then perturbed to pass sufficiently close to the black hole's
Schwarzschild radius to be captured. This model runs into difficulties
with the so-called So\l tan argument; since the SMBHs are fed mostly
dark matter rather than gas, there is no associated radiation. (2)
Direct capture of stars on high eccentricity orbits by the SMBH
~\cite{ZHR02,MP04}. This model has a similar problem because black
holes more massive than $\gsim 10^8~\msun$ do not tidally disrupt
stars, so there is again no radiative output associated with the black
hole growth. (3) Stellar captures by the accretion disk feeding the
hole \cite{MK05}.  

Solving the puzzle of the origin of the $\mh-\sigma$ relation will
have important implications for high-$z$ SMBHs: in particular, it will
generally determine how the relation evolves with redshift.

\subsection{The Future}
\label{sec:future}

In this section, we briefly summarize the possibility of probing the
continuum and line emission from AGN beyond the current redshift
horizon of $z\sim 6$.  This discussion is necessarily based on models
for how the BH population evolves at $z>6$.  Such models can be
constructed by assuming that SMBHs populate dark matter halos, e.g.,
in accordance with the locally measured $\mh-\sigma$ relation (or an
extrapolation of the relation to higher redshifts).  The relation
appears to hold, at least to within a factor of a few, for $z\sim 3$
quasars (this is based on using the H$\beta$/OIII lines as proxies for
black hole mass and $\sigma$, respectively; e.g., \cite{Shields+03}),
and also at $z\sim 6$ (see \S~\ref{zoltansubsec:sdss}).  No doubt
the observational constraints will improve as both black hole masses
and velocity dispersions are measured in larger samples of distant
quasars.  Correspondingly, extrapolations to high redshifts will be
more reliable as the feedback processes that regulate black hole
growth are better understood.  Here we summarize predictions from the
simplest models.

\subsubsection{Broadband Detections}

Predictions for the number counts of high redshift AGN have been made
using simple semi-analytic models for the near-infrared \cite{HL98b}
and in the soft X-rays \cite{HL99}.  In these early models, the
quasar black hole was assumed to have a fixed fraction $\sim 10^{-4}$
of the halo mass, shine at the Eddington luminosity, and have a duty
cycle of bright activity of $t_q\sim 10^6$~years.

In such models, the surface density of sources is very high in the
optical/near-infrared bands, even at $z\sim 10$.  For example, in the
$1-5\mu$m band, the $\sim 1$nJy sensitivity of the {\em James Webb
  Space Telescope (JWST)} will allow the detection of an $\sim
10^5~\msun$ black hole at $z=10$ (provided that the black hole shines
at the Eddington limit with a standard template
spectrum~\cite{Elvis+94}).  Surface densities as high as several
sources per square arcminute are predicted at this threshold from
$z~\gsim 5$, with most of these sources at $z~\gsim 10$ \cite{HL98a}.
We note, however, that these predictions are very sensitive to the
assumed duty cycle of bright activity.  For example, for $t_q\sim
10^7$~years, or $\mh \propto M_{\rm halo}^{5/3}$, the $z\sim 10$
counts can be smaller by a factor of 10-100 (depending on what
redshift--dependence is assumed for the above scaling relation between
black hole and halo mass at high redshift; see
\cite{HL98b,HNR98,WL03a} for related discussion).  It would also be
interesting to detect the host galaxies of ultrahigh redshift AGN,
which should be feasible with {\em JWST}'s sensitivity.  If the
galaxies occupy a fair fraction ($\sim 5\%$) of the virial radius of
their host halos, then a large fraction ($\gsim 50\%$) of them can
potentially be resolved with {\em JWST}'s planned angular resolution
of $\sim 0.06''$ \cite{HL98a,BL00}.  The Large Synoptic Survey
Telescope (LSST\footnote{www.lsst.org}), with a planned capability of
going $\sim5$ magnitudes deeper than SDSS in a $\sim 3$ times larger
solid angle, will be an ideal instrument for studying high-redshift
quasars in the optical/near-infrared.

In the soft X-rays, the $0.5-2$~keV flux of $2.5\times 10^{-17}~{\rm
  ergs~cm^{-2}~s^{-1}}$ reached in a 2~Ms exposure of CDF-North
\cite{Alexander+03} corresponds to a larger ($\sim 2\times
10^7~\msun$; see Figure 1 in \cite{HL99}) black hole at $z=10$, but
nevertheless, thousands of sources are predicted at $z\gsim 5$ per
square degree, and tens per square degree at $z>10$. This would imply
that tens of $z>5$ sources should have been detectable already in the
CDFs, whereas only a handful of potential candidates, and no confirmed
sources, have been found. In revised models with longer quasar
lifetimes and thus a steeper scaling of $\mh$ with $M_{\rm halo}$,
these numbers can be sharply decreased \cite{HL98a,HNR98}, which can
bring the expected counts into agreement with current non-detections
\cite{WL03a}.

The radio sensitivity of the extended Very Large Array and other
forthcoming instruments (e.g., Allen Telescope Array and Square
Kilometer Array) is also promising for detecting AGN beyond $z\sim
6$. Using the updated scaling of black hole mass with halo mass and
redshift~\cite{WL03a} and assuming the same radio-loud fraction ($\sim
10\%$) as at lower redshifts, a simple model predicts that $\sim $ten
$10\mu$Jy sources per square degree should be detectable at $1-10$~GHz
\cite{HQB04}. The identification of these quasars is a challenge, but
should, in principle, be feasible with deep optical/IR observations.
To date, only two such radio-selected quasars have been identified (in
optical follow-ups of sources in the FIRST and VLA radio
catalogs;~\cite{McGreer+06,Zeimann+11}), which falls below the
expectations from the simple model by a factor of several.

In addition to direct detection of AGN at very high redshifts, it may
also be possible to detect lower mass seed black holes at comparable
redshifts (or higher).  In particular, a plausible model for gamma-ray
bursts (GRBs) invokes accretion onto a newly formed $\sim 10~\msun$
black hole (the collapsar model; e.g., \cite{Woosley93}).  {\it Swift}
has now detected four bursts beyond $z>6$: GRB090429B at $z=9.4$
\cite{Cucchiara+11}, GRB090423 at $z=8.2$
\cite{Tanvir+09,Salvaterra+09}, GRB080913 at $z=6.7$
\cite{Greiner+09}, and GRB050904 at $z=6.3$
\cite{Kawai+06,Haislip+06}, for which the afterglow emission has also
been measurable; such afterglows should remain detectable in the
infrared out to $z\sim 20$ \cite{LR00,CL00}.  Their detection and the
characterization of their spectrum and light-curve would open up a new
probe of black hole formation and evolution at high redshifts.

In summary, model predictions for the continuum emission of $z>6$ AGN
are very sensitive to how one extrapolates the $\mh-M_{\rm halo}$
relation to $z\gsim 6$.  However, this should be viewed as ``good
news'': (1) large numbers of detectable AGN at these redshifts are
certainly possible, and (2) their detection will put strong
constraints on models for the origin and evolution of the black hole
population.

\subsubsection{Emission Line Measurements}

The strongest recombination lines of H and He from $5<z<20$ AGN will
fall in the near-infrared bands of {\em JWST} and could be bright
enough to be detectable.  Specific predictions have been made for the
source counts in the H$\alpha$ emission line \cite{Oh01} and for the
three strongest HeII lines \cite{OHR01,TGS01}.  The key assumption is
that most of the ionizing radiation produced by the miniquasars is
processed into such recombination lines (rather than escaping into the
IGM).  Under this assumption, the lines are detectable for a fiducial
$10^5~\msun$ miniquasar at $z=10$.  The Ly$\alpha$ line is more
susceptible to absorption by neutral hydrogen in the IGM near the
source but should be detectable for bright sources that are surrounded
by a large enough HII region so that Ly$\alpha$ photons shift out of
resonance before hitting the neutral IGM \cite{CH00}.  If the Lyman
$\alpha$ emission is scattered off expanding shells of material (as
expected from galactic winds), this will further shift the photons
away from resonance, and make the emission line more
detectable~\cite{DW10}.  We also note that in the ``trapped'' Lyman
$\alpha$ model for direct collapse mentioned in
\S~\ref{sec:directcollapse}, the Lyman $\alpha$ emission ultimately
emerging from the collapsing halo (before forming the SMBH) appears
detectable with {\it JWST}, as well -- as a low-surface brightness
diffuse blob~\cite{Latif+11b}.

The simultaneous detection of H and He lines would be especially
significant.  As already argued above, the hardness of the ionizing
continuum from the first sources of ultraviolet radiation plays a
crucial role in the reionization of the IGM. It would therefore be
very interesting to directly measure the ionizing continuum of any
$z>6$ source.  While this may be feasible at X-ray energies for
exceptionally bright sources, the absorption by neutral gas within the
source and in the intervening IGM will render the ionizing continuum
of high redshift sources inaccessible to direct observation out to
$1\mu$m.  This is a problem if the ionizing sources are black holes
with $M<10^8~\msun$ at $z\sim 10$ (easily detectable at wavelengths
red-ward of redshifted Ly$\alpha$ in the near-infrared by {\em JWST},
but too faint to see in X-rays).  The comparison of H$\alpha$ and HeII
line strengths can be used to infer the ratio of HeII to HI ionizing
photons, $Q=\dot{N}_{\rm ion}^{\rm HeII}/\dot{N}_{\rm ion}^{\rm HI}$.
A measurement of this ratio would shed light on the nature of the
first luminous sources, and, in particular, it could reveal if the
source has a soft (stellar) or hard (AGN-like) spectrum.  Note that
this technique has already been successfully applied to constrain the
spectra of sources in several nearby extragalactic HII
regions~\cite{Garnett+91}.  Lyman break galaxies at $z\approx 3$ also
appear to have unusually strong (for a normal stellar population)
He1640 emission line; however, the lack of X-rays rule out an AGN
explanation (the observation could be explained instead by the
presence of PopIII stars in these galaxies \cite{JH06} or by an
unusual abundant population of Wolf-Rayet stars \cite{BPC08} that can
produce the He1640 line while avoiding an overproduction of metal
lines).

Provided the gas in the high redshift AGN is enriched to near-solar
levels, several molecular lines may be visible. In fact, CO has
already been detected in the hosts of the most distant quasars
\cite{Walter+03,Wang+10,Wang+11}.  The detectability of CO for high
redshift sources in general has been considered in simple theoretical
models \cite{SS97}.  If AGN activity is accompanied by a star
formation rate of $\gsim 30~\msun/$~yr, the CO lines are detectable at
all redshifts $z=5-30$ by the Millimeter Array (the redshift
independent sensitivity is due to the increasing CMB temperature with
redshift), while the Atacama Large Millimeter Array (ALMA) could
reveal even fainter CO emission, and other C and O lines in emission,
providing spatially resolved images \cite{SSK10}.  The detection of
these molecular lines will provide valuable information on the stellar
content and gas kinematics near the AGN.

\subsubsection{Gravitational Waves}

The most direct observational constraints on the SMBH assembly at $z>6$, with
especially clear distinctions between the ``stellar--mass seed'' and
``direct-collapse'' scenarios, may come from detecting the
gravitational waves produced during the SMBH mergers.  The {\it Laser
Interferometer Space Antenna} ({\it LISA}) is expected to be able to
detect mergers of SMBHs in the mass range $\sim (10^4$--$10^7)\,{\rm
M_\odot}/(1+z)$ with high S/N out to $z\sim 30$ \cite{Baker+07}.
Binary spins and BH masses is expected to be measured with high
precision up to $z\sim 10$ \cite{Vecchio04}, especially if spin
precession \cite{LH06} and higher-order harmonics of the waveform
\cite{McWilliams+10} are included in the analysis.\footnote{As this
paper was being written, NASA announced a decision to withdraw from
the LISA experiment. The European Space Agency is continuing to
consider a redesigned version of LISA, with a smaller budget, and a
launch date of approximately 2021-2022. Given the very high S/N ratios
forecast for the original version of LISA, the redesigned
``LISA-lite'' mission should still be able to detect low-mass SMBHs
out to high redshifts.}  Many authors have computed the expected {\it
LISA} event rate from high--redshift SMBH merger population models in
a range of plausible models. The published estimates
(\cite{MHN01,WL03b,Sesana+04,Sesana+05,Islam+04,Sesana+07, LFH08}; see
a review in \cite{Arun+09}), even at lower redshifts, vary by orders
of magnitude, from $\sim 1$ to as high as $\sim 10^4$ yr$^{-1}$; there
is a large range even among models that are explicitly calibrated to
fit the evolution of the quasar luminosity function \cite{LFH08}.

Closest to the present context of the growth of SMBHs at $z>6$ are the
Monte-Carlo merger tree models in ref.~\cite{TH09} (discussed in
\S~\ref{sec:stellarseedexample} above).  These models coupled the
merger trees with the orbits of oscillating kicked BHs, to predict
detection rates for {\it LISA}.  They have surveyed a wide range of
candidate assembly models, including those with rare, massive seeds,
or through ultra--early production of numerous Pop-III remnant seeds.
As mentioned above, in the latter model, seed BHs need to stop forming
below a redshift $z_{\rm cut}\sim 20$, in order to avoid overproducing
$10^6~{\rm M_\odot}$ BHs.

The simplest SMBH assembly scenarios, which have constant accretion
rates, but in which BH seed formation stops abruptly at some redshift,
and which meet constraints at both the high--mass and low--mass end of
the $z=6$ SMBH mass function, predict negligibly low {\it LISA} event
rates.  The reason for this pessimistic conclusion is as follows: in
these models, the BHs that grow into the most massive,
highest-redshift quasar-SMBHs accrete at the same (exponential) rate
as all the other BHs, typically resulting in a vast overproduction of
massive ($m\sim 10^{6}\Msol$) holes.  In order to offset this
overproduction, seeds must be made very rare, and this diminishes the
{\it LISA} rates.  It is difficult to envision a scenario for high
($\gsim 10$ per year per unit redshift) detection rates unless a vast
number of SMBHs in the $10^{5-7}\Msol$ range lurk in the universe at
all redshifts, which the current electromagnetic surveys have missed.

A different class of models, which successfully build the $z\sim 6$
quasar BHs, are those in which the SMBH masses are self--regulated by
internal feedback, to always maintain the $\mh-\sigma$ relation. These
models can evade this constraint, and produce {\it LISA} rates as high
as $30$ yr$^{-1}$.  The key difference in these models with higher
{\it LISA} rates is that the SMBH growth is driven by a large number
of seed BHs and far lower gas accretion rates than those required in
the constant-accretion models.  The majority of the {\it LISA} events
occur at $z\approx 6$ and in the low end ($10^3-10^4$~${\rm M_\odot}$)
of {\it LISA}'s mass range for detection.

Also, for these models, the ejected BH mass density can exceed that of
the galactic BH population at $z=6$.  Most ejected holes are expected
to have low masses (still similar to the original seed mass), but an
ejected BH can be as massive as $\sim10^{8}\Msol$ if large recoil
velocities are allowed (e.g. if spins are not always aligned with the
orbital angular momentum of the binary).

Using similar ``merger tree'' massive black hole formation models,
\cite{Sesana+07} analyzed the predicted mass-- and
redshift--distribution of LISA events.  These models have input
assumptions similar to the ``$\mh-\sigma$`` models in \cite{TH09}, but
with varying initial seed masses. These predict a handful of
detectable events at $z>10$. The raw total event rates in the two
models are very similar. However, the mass-distribution of events is
different (low-mass mergers are missing in the 'heavy seed'
model). Another key diagnostic between the 'heavy' and 'stellar-mass'
seed models is the mass ratio of the BHs in these detectable events:
while the former models predict near-equal mass mergers, in the latter
case, one of the merger partners typically have time to grow,
resulting in typical mass ratios of $q=0.1-0.2$.

It is worthwhile to note that essentially all of the work on the
gravity wave signal from black hole-black hole in-spiral has assumed
efficient (nearly instantaneous) mergers.  Stellar-scattering and gas
can help drive BHs together on large scales, which can affect
detections of the most massive nearby SMBHs by Pulsar Timing Arrays
\cite{KS11} and extreme mass ratio inspiral events by {\em LISA}
\cite{Yunes+11}.  However, SMBH-SMBH coalescences, when they are in
{\em LISA}'s frequency window, are well within the rapidly merging
purely gravitational wave-driven regime, at least if the circumbinary
gas forms a thin disk \cite{HKM09}.

\section{Conclusions}
\label{sec:conclude}

In this review, we have summarized theoretical ideas and observational
constraints on how massive black holes form at the centers of the
earliest protogalaxies, and how such black holes grow via accretion and
mergers to give rise to the observed population of black holes at
$z\gsim 6$ and in the local and moderate redshift universe.  As this
review shows, this remains a poorly understood but rich and important
problem.  Perhaps the most direct way of probing the role of mergers
in black hole assembly and evolution at $z>10$ is via their gravity
wave signatures, which will hopefully be feasible with the redesigned
version of LISA being considered by ESA.

In addition to being of intrinsic interest for understanding the AGN
phenomena, sources of gravity waves, etc., there is strong evidence
that the formation and evolution of black holes is coupled to the
formation and evolution of the host galaxy in which the black hole
resides (e.g., the $\mh-\sigma$ relation), and thus to the
cosmological formation of nonlinear dark matter structures (i.e., the
dark halos surrounding these galaxies).  We anticipate that this will
remain a growth area of research in the coming years, with continued
rapid progress on both the observational and theoretical fronts.

\begin{acknowledgement}
  The author wishes to thank the editors of this volume for their
  patience during the preparation of this article.  I would also like
  to thank my recent collaborators, especially Taka Tanaka, Cien
  Shang, Mark Dijkstra and Greg Bryan, whose work was especially
  emphasized here, and Eliot Quataert for permission to draw on
  material in our earlier joint review. The work described here was
  supported in part by the NSF, NASA, and by the Pol\'anyi Program of
  the Hungarian National Office for Research and Technology (NKTH).
\end{acknowledgement}

\bibliographystyle{plain}

\end{document}